\documentclass[twocolumn,showpacs,preprintnumbers,amsmath,amssymb,superscriptaddress,aps,prc]{revtex4-1}
\usepackage{graphicx}
\usepackage{xcolor}
\usepackage{subfigure}
\usepackage{amssymb}
\usepackage{graphicx}
\usepackage{subfigure}
\usepackage{amsmath}
\usepackage[mathscr]{eucal}

\begin{document}

\preprint{APS/123-QED}

\title{Improved Wong and classical approximations, and reduction of fusion data}

\author{L.F. Canto}
	\email{canto@if.ufrj.br}
	\affiliation{Instituto de F\'{\i}sica, Universidade Federal do Rio de Janeiro, 
 CP 68528, 21941-972, Rio de Janeiro, RJ, Brazil}

 \author{V. A. B. Zagatto} 
	\email{vzagatto@id.uff.br}	
	\affiliation{Instituto de F\'{\i}sica, Universidade Federal Fluminense, 
 Av. Litoranea s/n, Gragoat\'{a}, Niter\'{o}i, R.J., 24210-340, Brazil}	

\author{J. Lubian} 
\email{jlubian@id.uff.br}
	\affiliation{Instituto de F\'{\i}sica, Universidade Federal Fluminense, 
 Av. Litoranea s/n, Gragoat\'{a}, Niter\'{o}i, R.J., 24210-340, Brazil}

\author{R. Donangelo}
	\email{donangel@fing.edu.uy}
	\affiliation{Instituto de F\'\i sica, Facultad de Ingenier\'\i a, 
 Universidad de la Rep\'ublica,
 Av. Julio Herrera y Reissig 565, 11300 Montevideo, Uruguay}

\begin{abstract}
We present an improved version of the Wong formula for heavy-ion fusion, 
where the parameters of the parabolic approximation of the Coulomb barrier 
are replaced by parameters of the $l$-dependent potential at an
effective partial-wave. A pocket formula for this $l$-dependence is given.
This version reproduces the fusion cross sections of quantum 
mechanical calculations very well, even when the original Wong formula is not valid. 
The same procedure is used to derive an improved expression for the 
classical fusion cross section, which is very accurate at above-barrier
energies.
Based on this classical expression, we propose a new method to reduce 
fusion data in this energy range. 
This method is used to perform a comparative study of complete fusion 
suppression in collisions of weakly bound projectiles. 
This study indicates that the suppression of complete fusion is 
essentially due to the action of nuclear breakup couplings.
\end{abstract}

\maketitle

%%%%%%%%%%%%%%%%%%%%%%%%%%%%%%%%%%%%%%%%%%%%%%%%%%%%%%%%%%%%%%%%%%%%%%%%%%%%%%%%%%

\section{Introduction}

 The dynamics of fusion reactions are highly complex, 
especially in collisions of projectiles with low breakup thresholds (lower than 3 or 4 MeV)~\cite{CGD06,KRA07,KAK09,CGD15,KGA16,CGL20}. Owing to 
the low binding energy, the projectile can break up as it approaches the target, giving rise to different fusion
processes. First, is the direct complete fusion (DCF), where the whole projectile fuses with the target. 
This is the usual fusion reaction, observed also in the collisions of tightly bound nuclei. In addition, the breakup of the projectile triggers two more fusion
processes: the incomplete fusion (ICF) and the sequential complete fusion (SCF).
The former occurs when at least one, but not all, projectile fragments fuse 
with the target. The latter takes place when all breakup fragments fuse 
sequentially with the target. 
Experimentally, SCF cannot be distinguished from DCF. The experimental cross section corresponds to complete fusion
(CF), the sum DCF + SCF. Besides, many experiments cannot distinguish CF from ICF. Then, the data
correspond to total fusion (TF), the sum of all fusion processes.\\

In collisions of weakly bound projectiles, the low breakup threshold affects the CF cross section in two
ways. Owing to the low binding energy, the projectile's density has a long tail, leading to a reduction of the 
Coulomb barrier. This is a static effect, which enhances CF at all collision energies. On the other hand, 
the low breakup threshold leads to strong breakup couplings, diverting an appreciable part of 
the incident flux into the breakup channel. This is a dynamic effect that hinders CF.  In this way, the CF data is the net result of the competition between the opposing static and
dynamic effects of the low binding energy of the projectile. Experimental 
and theoretical studies indicate that the CF cross section
is enhanced at sub-barrier energies and suppressed above the Coulomb 
barrier~\cite{CGD06,KRA07,KAK09,CGD15,KGA16,CGL20}. However, the present
understanding of CF reactions is not entirely satisfactory. New experimental and theoretical
studies are called for. \\

There are different methods to assess the influence of the low breakup threshold of the 
projectile on the CF cross section. The first is to compare the CF data to predictions 
of theoretical models that do not consider the projectile's low binding. A
standard procedure is to compare them to fusion cross sections of one-channel 
calculations with a standard real potential and short-range absorption. 
Usually, one adopts implementations of the double-folding model 
with standard nuclear densities, like the  S\~ao Paulo (SPP)~\cite{CPH97,CCG02} or the Aky\"uz-Winther (AW)~\cite{AkW81} potential. 
A more straightforward approach is to compare the data to predictions of semiclassical or classical 
approximations to the quantum mechanical (QM) treatment, like the barrier penetration 
model (BPM), the Wong formula~\cite{Won73}, or even the classical fusion cross section. 
However, one should ensure that the approximate model for the theoretical cross 
section is valid under the conditions of the experiment. Then, the differences between the experimental
and the theoretical cross sections are assumed to arrive from the low binding energy of the projectile.\\

The other possibility is to compare CF data of the weakly bound system to fusion data of similar tightly bound systems.
However, direct comparisons of the data would be meaningless since they are strongly dependent on the charges and masses 
of the collision partners. Thus, submitting the data to some reduction procedure that eliminates the
influence of such trivial factors is necessary. Several reduction procedures have been proposed (for a review, see \cite{CMG15}).
A very efficient one is the fusion function (FF) reduction method, which is based on the Wong formula~\cite{Won73}. 
In an ideal situation where the fusion cross section is unaffected by the collision partners' intrinsic structure, and the Wong formula approximates the QM cross section, the reduced
data is well described by a universal fusion function. Then, this universal function is used as a benchmark to assess
the importance of channel coupling effects in the fusion data. However, the Wong formula is not a good approximation
to the QM cross section for light systems at sub-barrier energies and at energies well above the 
barrier~\cite{CGL09a,CGL09b,CGL09c}.\\

In the present paper, we discuss the limitations of the Wong formula and the classical expression for the fusion
cross section. We show that their validity can be extended if one replaces the s-wave barrier parameters with effective barrier parameters obtained by proper angular-momentum averages. Then, we propose a new reduction 
method for the above-barrier fusion data based on this improved version of the classical cross section. This method is used 
to study CF data of weakly bound systems.\\

The paper is organized as follows. In Sec.~\ref{theory}, we discuss the use of complex interactions in potential scattering
to simulate the effects of fusion and total reaction in the scattering wave function. 
In sec.~\ref{Clas-Wong}, we introduce the classical and Wong approximations to the QM fusion cross section. We then
discuss the validity of these approximations and the classical limit of the Wong formula.
In Sec.~\ref{improv-Wong}, we develop new versions of Wong and the classical approximations and then propose 
a new reduction method based on the latter. 
In Sec.~\ref{app}, we use the new reduction method to perform a comparative study of CF data in collisions 
of $^6$Li, $^7$Li, $^9$Be, and $^6$He on several targets. 
Finally, in Sec.~\ref{concl}, we give the main conclusions of the present paper.

%%%%%%%%%%%%%%%%%%%%%%%%%%%%%%%%%%%%%%%%%%%%%%%%%%%%%%%%%%%%%%%%%%%%%%%%%%%%%%%%%%%%%

\section{The fusion and the reaction cross sections in potential scattering}\label{theory}

%%%%%%%%%%%%%%%%%%%%%%%%%%%%%%%%%%%%%%%%%%%%%%%%%%%%%%%%%%%%%%%%%%%%%%%%%%%%%%%%%%%%%

In an idealized situation where the intrinsic degrees of freedom of the collision 
partners do not affect the collision, the elastic cross section can be described by 
potential scattering. 
In this approach, the colliding nuclei are treated as point particles, interacting 
through Coulomb and nuclear forces. 
The scattering wave function is expanded in partial waves and the resulting radial 
wave functions are calculated by solving the radial equation with the $l$-dependent 
potential
\begin{equation}
V_l(r) = V_{\rm N}(r) + V_{\rm C}(r) + \frac{\hbar^2\ l(l+1)}{2\mu r^2}.
\label{Vl}
\end{equation}
Above, $V_{\rm N}(r)$ is the nuclear interaction between the projectile and the target, where $r$ is the distance between them. 
It can be obtained by integrating the densities multiplied by a properly chosen 
nucleon-nucleon interaction (folding model). 
The  S\~ao Paulo (SPP)~\cite{CPH97,CCG02} and the Aky\"uz-Winther (AW)~\cite{AkW81} 
potentials are implementations of the folding model, using different approximations.  
They are widely used in the literature.\\

The Coulomb potential is usually approximated as
\begin{eqnarray}
V_{\rm C}(r) &=& \frac{Z_{\rm P} Z_{\rm T} e^2}{2R_{\rm C}}  \ 
\left( 3 - \frac{r^2}{R_{\rm C}^2}\right); \ \ \ {\rm for }\ \ 
r<R_{\rm C}\nonumber\\
&=& \frac{Z_{\rm P} Z_{\rm T} e^2}{r}; \qquad\qquad\ \qquad\  {\rm for }\ \ 
r \ge R_{\rm C}.
\label{VC}
\end{eqnarray}
Above, $R_{\rm C}$ is the Coulomb radius, corresponding to the sum of the radii 
of the collision partners, and $Z_{\rm P}$ and $Z_{\rm T}$ are  
the atomic numbers of the projectile and the target, respectively.\\

The third term in Eq.~(\ref{Vl}) is the centrifugal potential, which accounts 
for the tangential kinetic energy in the radial equation. It
is inversely proportional to the reduced mass, $\mu$, and grows quadratically
with the angular momentum ($\sim l^2$ for large $l$).\\

The attractive nuclear potential is very strong but has a short range. 
On the other hand, the repulsive Coulomb and centrifugal terms are weaker 
but have longer ranges. The competition between the attractive and repulsive 
terms leads to a barrier in the potential $V_l(r)$, located at $R_l$ and 
with height $B_l$. \\

However, actual nucleus-nucleus collisions are always affected by intrinsic 
degrees of freedom, which are coupled with the coordinate $r$. 
Then, a fraction of the incident flux is diverted into nonelastic channels 
along the collision. 
In potential scattering, this effect is simulated by the addition of an 
attractive imaginary part to the nuclear potential, namely
\[
V_{\rm N}(r)\,\longrightarrow\, U_{\rm N}(r) = V_{\rm N}(r) + i W(r).
\]
In this way, the S-matrix loses its unitarity, giving rise to an absorption 
cross section. 
For a collision with energy $E$ (wave number $k$), it is given by
\begin{equation}
\sigma_{\rm abs} (E)= \frac{\pi}{k^2} \sum (2l+1)\ P_{\rm abs}(l,E).
\label{sigabs}
\end{equation}
Above, $P_{\rm abs}(l,E)$ is the absorption probability,
\begin{equation}
P_{\rm abs}(l,E) = 1 - \left|S_l(E)\right|^2,
\label{Pabs}
\end{equation}
where, $S_l(E)$ is the $l$-th component of the S-matrix. \\

The characteristics of the imaginary potential depend on the nuclear reactions 
whose effects it is simulating. 
This point is discussed below.

%%%%%%%%%%%%%%%%%%%%%%%%%%%%%%%%%%%%%%%%%%%%%%%%%%%%%%%%%%%%%%%%%%%%%%%%%%%%%%%%

\subsection{The total reaction cross section}

%%%%%%%%%%%%%%%%%%%%%%%%%%%%%%%%%%%%%%%%%%%%%%%%%%%%%%%%%%%%%%%%%%%%%%%%%%%%%%%%

The elastic cross section is influenced by fusion and nonelastic processes. 
The former only takes place when the densities of the collision partners overlap
strongly. 
On the other hand, the latter are dominant in grazing collisions, where the 
distance of closest approach is larger than the barrier radius. 
Then, if one wants to evaluate the total reaction cross-section, which has 
contributions of both reaction mechanisms, the imaginary potential must 
be very strong in the inner region of the barrier but also act in the neighborhood 
of the barrier radius. 
Then, the range of $W(r)$ must be similar to that of the real part of the nuclear 
potential. \\

Frequently, the imaginary potential used in calculations of the total reaction cross section is represented by a Woods-Saxon (WS) function,
\begin{equation}
W_{\rm R}(r) = -\ \frac{W_0}{
1 + \exp\left[ \left( r - R_{\rm w} \right)/a_{\rm w} \right]} .
\label{W-SWR}
\end{equation}
%
%reaching the barrier region. 
Above,
\begin{equation}
R_{\rm w} = r_{\rm w}\,\left(A_{\rm P}^{1/3} + A_{\rm T}^{1/3}\right),
\end{equation}
where $A_{\rm P}$ and $A_{\rm T}$ stand respectively for the mass numbers of the 
projectile and the target. 
Adopting parameters of the order $W_0 \simeq 50$~MeV, $r_{\rm w} \sim 1.2$~fm, 
and $a_{\rm w} \sim 0.6$~fm, the imaginary potential has the desired behavior.\\

Another possibility is to take the imaginary potential proportional to the real one. 
That is, one writes
\begin{equation}
U_{\rm N}(r) = (1 +\beta\, i)\,V_{\rm N}(r),
\label{WV}
\end{equation}
where $\beta$ is a constant, usually close to one. 
Gasques {\it et al.}~\cite{GCG06} used the SPP together with imaginary potentials 
of this kind to successfully analyze experimental scattering and  total reaction 
data of a large number of systems.

%%%%%%%%%%%%%%%%%%%%%%%%%%%%%%%%%%%%%%%%%%%%%%%%%%%%%%%%%%%%%%%%%%%%%%%%

\subsection{The fusion cross section}

%%%%%%%%%%%%%%%%%%%%%%%%%%%%%%%%%%%%%%%%%%%%%%%%%%%%%%%%%%%%%%%%%%%%%%%%

An imaginary potential to simulate the effects of fusion in potential scattering 
must lead to total absorption in the inner region of the barrier and be negligible 
elsewhere. 
This behavior is  guaranteed by a WS function like that of Eq.~(\ref{W-SWR})
with a short range, i.e., with parameters like \\
\begin{equation}
W_0 \simeq 50\, {\rm MeV}, \ \ r_{\rm w} \sim 1.0\, {\rm fm}\ \ {\rm and}\ \ 
a_{\rm w} \sim 0.2\, {\rm fm}.
\label{WS-short}
\end{equation}

 Denoting by  $S_l^{\rm F}(E)$ and $ P_{{\rm F}}(l,E)$ the partial-wave components 
 of the S-matrix and the corresponding fusion probability in a calculation with 
 the imaginary potential of Eqs.~(\ref{W-SWR}) and (\ref{WS-short}), the fusion 
 cross section can be written as
\begin{equation}
\sigma_{\rm F} (E)= \frac{\pi}{k^2} \sum_{l=0}^{\infty} (2l+1)\ P_{\rm F}(l,E),
\label{sigF0}
\end{equation}
with 
\begin{equation}
P_{{\rm F}}(l,E) = 1 -\left| S_l^{\rm F}(E)\right|^2.
\label{PF}
\end{equation}

 We remark that similar results can be obtained with a real nuclear potential 
 but adopting ingoing wave boundary  conditions~\cite{Raw66}. 
This approach is used in the CCFULL computer code~\cite{HRK99}, frequently used 
in coupled channel calculations.\\

The above discussion suggests that the fusion probability in Eq.~(\ref{sigF0}) 
is equivalent to the probability of the incident wave reaching
the strong absorption region. 
In this way, the fusion cross-section could be approximated by the cross-section 
of the barrier penetration model (BPM),
\begin{equation}
\sigma_{\rm BPM}(E) =  \frac{\pi}{k^2} \sum_{l=0}^{\infty} (2l+1)\ T(l,E).
\label{sigF1}
\end{equation}
Above, $T(l,E)$ is the transmission coefficient for the system to go through 
the barrier of the potential $V_l(r)$ in a collision with energy $E$. 
It is calculated by Kemble's version~\cite{Kem35} of the WKB approximation, 
through the expression
\begin{equation}
T(l,E) \equiv T_{\rm WKB}(l,E) = \frac{1}{1+\exp\left[2\,\Phi(l,E) \right]}.
\label{TK}
\end{equation}
At sub-barrier energies, $\Phi(l,E)$ is the integral of the imaginary 
wave number between the classical turning points. 
That is,
\begin{equation}
\Phi(l,E) = \int_{r_i}^{r_e} \kappa_l(r)\ dr,
\label{PhiWKB}
\end{equation}
where 
\begin{equation}
\kappa_l(r) = \sqrt{
\frac{2\mu}{\hbar^2}\,\Big[ V_l(r) - E\Big].
}
\label{kappa}
\end{equation}
The internal ($r_i$) and external ($r_e$) turning points are determined by 
the conditions\\
\begin{equation}
V_l\left( r_i\right) = E,\, r_i < R_l \ \ \ {\rm and}\ \ V_l
\left( r_e\right) = E,\, r_e \ge R_l .
\label{tp}
\end{equation}

However, this prescription cannot be used above the barrier, with no real turning points. 
A solution to this problem was found by Hill and Wheeler~\cite{HiW53}. 
First, they approximated the potential barrier by the parabola 
\begin{equation}
V(r) \simeq B_l - \frac{1}{2} \mu \omega_l^2 \left(r-R_l\right)^2,
\label{Vl-parab}
\end{equation}
where $B_l,R_l$ and $\hbar\omega_l$ are the height, radius and barrier 
curvature parameters. 
Of particular interest are the parameters for the Coulomb barrier, denoted by
\begin{equation}
V_{\rm B} \equiv B_{l=0},\qquad  R_{\rm B} \equiv R_{l=0},\;\;\; {\rm and}\;\;\; 
\hbar\omega \equiv \hbar\omega_{l=0}.
\label{VB-RB-hw}
\end{equation}
Next, they carried out the analytical continuation of the collision coordinate 
onto the complex plane and found the complex turning points 
(see also Ref.~\cite{TCH17}). 
In this way, the WKB integral of Eq.~(\ref{PhiWKB}) could be evaluated analytically, 
and they found the result\\
\begin{equation}
\Phi(l,E) = \frac{\pi}{\hbar\omega} \big(B_l - E \big).
\label{PhiHW}
\end{equation}
\begin{table}%[h!]
\centering
\caption{Barrier parameters of the S\~ao Paulo potential for the systems discussed 
in the text. 
The barrier radii are expressed in fm and height and curvature
parameters are given in MeV.}
\vspace{0.5cm}
\begin{tabular}{lcccc}
\hline 
System                        \qquad\qquad\ \  \               &  
$R_{\rm B}$ \qquad \   &\qquad\  $V_{\rm B}$  &\qquad\   $\hbar\omega$ \\ 
\hline
$^7$Li + $^{27}$Al      \qquad\qquad\ \  \               &    8.5               
\qquad\ &  \qquad\ 6.1              &  \qquad\ 2.4                \\
%\hline  
$^7$Li + $^{209}$Bi      \qquad\qquad\ \ \              &     11.4            
\qquad\ & \qquad\ 29.4             &   \qquad\ 4.1               \\
%\hline  
$^{24}$Mg + $^{138}$Ba  \qquad\qquad \ \ \        &     11.1            
\qquad\ & \qquad\ 81.6            &  \qquad\ 3.9               \\
%\hline  
%
\hline
\end{tabular}
\label{barrier parameters}
\end{table}
\begin{figure*}%[h!]
\begin{center}
\includegraphics*[width=14 cm]{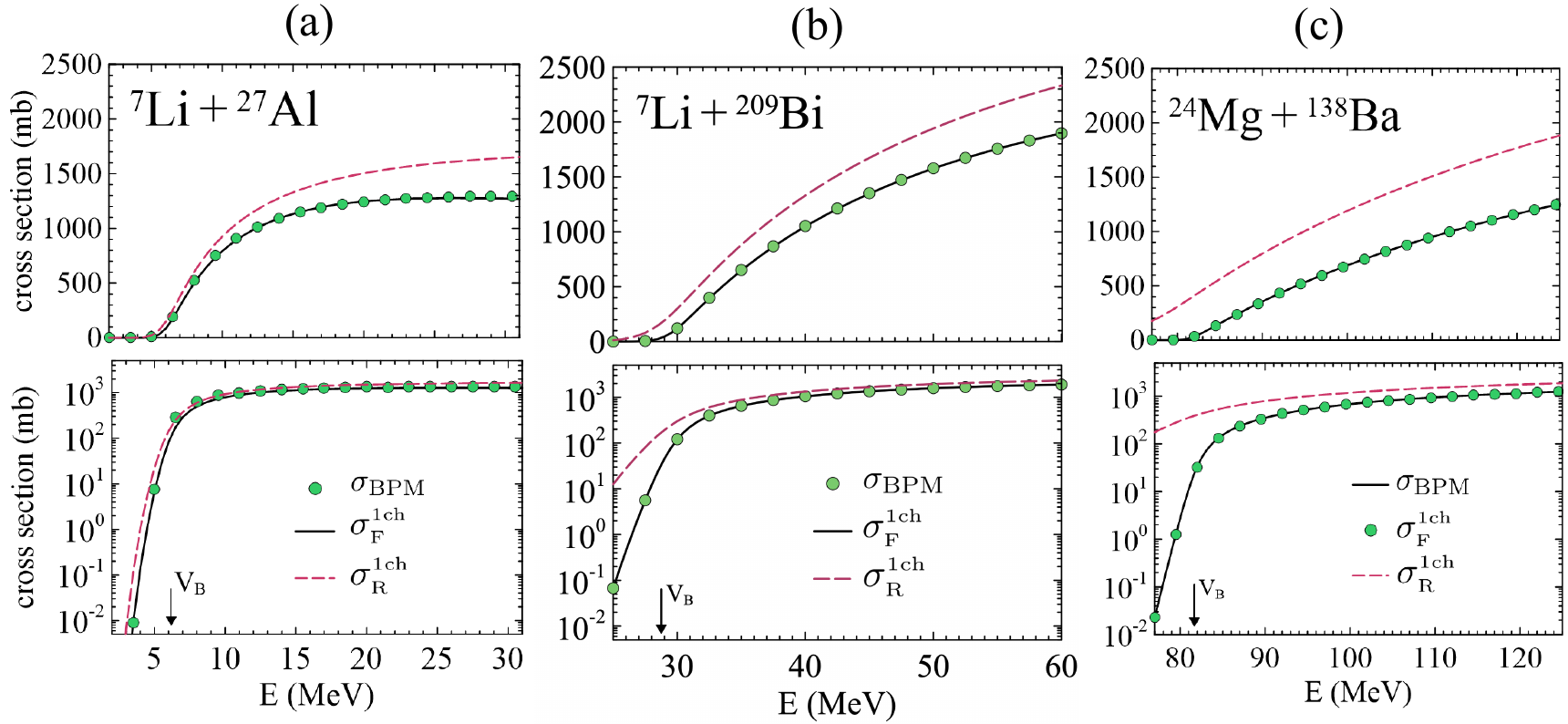}
\end{center}
\caption{Comparison of BPM fusion cross sections, $\sigma_{\rm BPM}$, with 
fusion cross sections of QM calculations with the SPP and short-range absorption, 
$\sigma_{\rm F}^{\rm 1ch}$, for three systems in different mass regions. 
For comparison, we also show cross-sections of QM calculations with the 
long-range imaginary potential of Eq.(\ref{WV}), $\sigma_{\rm R}^{\rm 1ch}$.}
\label{BPM-vs-QM}
\end{figure*}
We compared BPM cross-sections to the ones obtained by solving the one-channel 
radial equations with a short-range imaginary potential. These cross sections are denoted by 
$\sigma_{\rm BPM}$ and $\sigma_{\rm F}^{\rm 1ch}$, respectively. We also compared
the total reaction cross sections, denoted by $\sigma_{\rm R}^{\rm 1ch}$.\\

We considered a light, $^7$Li\,+\,$^{27}$Al, an intermediate, $^7$Li\,+\,$^{209}$Bi, 
and a heavy system, $^{24}$Mg\,+\,$^{138}$Ba. Here, and throughout the present work, 
we adopt the SPP for the real part of the nuclear potential. The barrier parameters of 
the parabolic expansion of $V_{l=0}(r)$ for the above-mentioned systems are listed 
in Table~\ref{barrier parameters}.
In the calculations of $\sigma_{\rm F}^{\rm 1ch}$, we used strong absorption imaginary
potentials with a short range. They were given by a WS function with the parameters of 
Eq.~(\ref{WS-short}). In the calculations of $\sigma_{\rm R}^{\rm 1ch}$, the imaginary 
potential was given by Eq.~(\ref{WV}), with $\beta = 0.78$~\cite{GCG06}. 
\\

The $\sigma_{\rm BPM}$, the $\sigma_{\rm F}^{\rm 1ch}$, and the $\sigma_{\rm R}^{\rm 1ch}$ 
cross sections for the $^7$Li\,+\,$^{27}$Al, $^7$Li\,+\,$^{209}$Bi, and $^{24}$Mg\,+\,$^{138}$Ba systems 
are shown in Fig.~\ref{BPM-vs-QM}. The calculations were performed at collision energies 
ranging from $\sim 4$~MeV below $V_{\rm B}$ to $\sim 30$ MeV above it. In each case, the 
results are presented in linear and logarithmic scales, appropriate for comparisons 
at the above-barrier and sub-barrier energies, respectively. We find that the 
$\sigma_{\rm F}^{\rm BPM}$ and the $\sigma_{\rm F}^{\rm 1ch}$ cross sections for the three 
systems are really very close. The corresponding curves can hardly be distinguished in the figures. 
On the other hand, the total reaction cross sections are much larger than the other two,
mainly at sub-barrier energies. 
This is not surprising since $\sigma_{\rm R}^{\rm 1ch}$ has contributions 
from absorption beyond the barrier radius, which remains relevant even at 
collision energies well below $V_{\rm B}$. 
Then, we stress that the BPM is a very good approximation to fusion  
but very different from the total reaction cross-section.\\

Since $\sigma_{\rm BPM}$ is practically identical to the QM fusion cross section, 
$\sigma_{\rm F}^{\rm 1ch}$, we henceforth consider the former as the benchmark
cross section to assess the accuracy of approximate expressions.

%%%%%%%%%%%%%%%%%%%%%%%%%%%%%%%%%%%%%%%%%%%%%%%%%%%%%%%%%%%%%%%%%%%%%%%%%%%%%%%%%%%%%%%%%

\section{The classical and the Wong approximations to the fusion cross section}
\label{Clas-Wong}

%%%%%%%%%%%%%%%%%%%%%%%%%%%%%%%%%%%%%%%%%%%%%%%%%%%%%%%%%%%%%%%%%%%%%%%%%%%%%%%%%%%%%%%%%

%%%%%%%%%%%%%%%%%%%%%%%%%%%%%%%%%%%%%%%%%%%%%%%%%%%%%%

\subsection{The classical approximation}

%%%%%%%%%%%%%%%%%%%%%%%%%%%%%%%%%%%%%%%%%%%%%%%%%%%%%%

At high enough collision energies (the meaning of high enough will be clarified later), 
one can use the classical approximation for the fusion cross-section. 
To derive it, one makes the following assumptions:
\begin{enumerate}

\item The sum of partial waves involves many $l$-values, so that the angular momentum 
can be treated as a  continuous variable, $\lambda$:
\begin{equation}
l\,\rightarrow\, \lambda = l + 1/2. 
\label{continuous l}
\end{equation}
In this way, the sum over partial-waves becomes an integral, namely

\begin{equation}
\sum_{l=0}^\infty (2l+1)\,\rightarrow\, \int_{\lambda = 1/2}^\infty 2\lambda \, d\lambda.
\label{sum-integral}
\end{equation}
Then, one replaces in Eq.~(\ref{Vl}): $l(l+1)=\lambda^2 - 1/4$, and one 
gets the $\lambda$-dependent potential
\begin{equation}
V_\lambda(r) = V_{\rm N}(r) + V_{\rm C}(r) + \frac{\hbar^2}{2\mu r^2}\, 
\left( \lambda^2-1/4 \right).
\label{Vlambda}
\end{equation}

\item Next, one neglects the $\lambda$-dependence of $R_\lambda$ and $\hbar\omega_\lambda$, and writes
\begin{equation}
R_\lambda = R_{\rm B};\ \ \hbar\omega_\lambda = \hbar\omega.
\label{l-indep-R}
\end{equation}
In this way, one gets the barrier heights
\begin{equation}
B_\lambda = V_{\rm B} + \frac{\hbar^2}{2\mu R_{\rm B}^2}\, \left( \lambda^2-1/4 \right).
\label{Bl-1}
\end{equation}
\item Tunneling effects are neglected, so that the transmission coefficient becomes
\begin{eqnarray}
T(\lambda,E) \simeq T_{\rm cl}(\lambda,E) & = & 1\  \ {\rm for}\ \ \lambda\le \lambda_{\rm g},\nonumber \\
                              & = & 0\  \ {\rm for}\ \  \lambda >  \lambda_{\rm g}.
\label{Tclass}
\end{eqnarray}
\end{enumerate}
Above, $\lambda_{\rm g}$ is the grazing angular momentum in a collision 
with energy $E$, given by the condition\\
\begin{equation}
B_{\lambda_{\rm g}} = E.
\label{lg}
\end{equation}
Using the explicit form of $B_\lambda$, within the approximation of 
Eq.~(\ref{Bl-1}), one gets the relation
\begin{equation}
 \lambda_{\rm g}^2 - \frac{1}{4}= \frac{2\mu\,R_{\rm B}^2}{\hbar^2}
 \left( E-V_{\rm B}\right).
\label{lambda_g}
\end{equation}
%
%\end{enumerate}

\bigskip

Within the above approximations, we can derive an analytical expression for 
the fusion cross section of Eq.~(\ref{sigF1}). 
One gets,
\begin{equation}
\sigma_{\rm cl}(E) =   \frac{\pi}{k^2}  \int_{\lambda = 1/2}^\infty 2\lambda\, 
T(\lambda,E) \ d\lambda ,
\label{sigF-cl-0}
\end{equation}
or, using Eq.~(\ref{Tclass}), 
\begin{equation}
\sigma_{\rm cl}(E) =   \frac{\pi}{k^2}  \int_{\lambda = 1/2}^{\lambda_{\rm g}} 
2\lambda \ d\lambda =
\frac{\pi}{k^2} \left( \lambda_{\rm g}^2 -\frac{1}{4} \right).
\label{sigF-cl1}
\end{equation}
Then, using Eq.~(\ref{lambda_g}), one gets the classical fusion cross section \\
\begin{eqnarray}
\sigma_{\rm cl}(E) &=&  \pi R_{\rm B}^2\ \left( 1 -\frac{V_{\rm B}}{E} \right),
\qquad {\rm for}\ E\ge V_{\rm B};\label{sigclass} \\
                   &=&  0 \qquad\qquad \qquad\qquad\ \ \ \ 
                   {\rm for}\ E < V_{\rm B}\nonumber.
\end{eqnarray}

The classical fusion cross section has a serious flaw: it vanishes at 
sub-barrier energies.

%%%%%%%%%%%%%%%%%%%%%%%%%%%%%%%%%%%%%%%%%%%%%%%%%%%%%%

\subsection{The Wong formula}

%%%%%%%%%%%%%%%%%%%%%%%%%%%%%%%%%%%%%%%%%%%%%%%%%%%%%%

 C.Y. Wong~\cite{Won73} derived an analytic expression for the fusion 
 cross section, which includes tunneling effects. 
 To get his formula, Wong made the same assumptions as in the derivation 
 of the classical cross section, except for the transmission coefficient of 
 Eq.~(\ref{Tclass}). 
 Instead, he used the Hill-Wheeler transmission coefficient,
\begin{equation}
T_{\rm HW}(l,E) = \frac{1}{1+\exp\left[\frac{2\pi}{\hbar\omega}\,
\left(V_{\rm B} - E\right) \right]},
\label{THW}
\end{equation}
below and above the barrier. 
Then, Eq.~(\ref{sigF-cl-0}) becomes 
\begin{equation}
\sigma_{\rm W}(E) =   \frac{\pi}{k^2}  \int_{\lambda = 1/2}^\infty \
\frac{2\lambda\, d\lambda}
{1+\exp\left[\frac{2\pi}{\hbar\omega}\,\left(V_{\rm B} - E\right) \right]}.
\label{sigF-W-0}
\end{equation}
The above integral can be evaluated analytically; the result is the 
Wong formula, 
\begin{equation}
\sigma_{\rm W} =\frac{\hbar\omega\,R_{\rm B}^2}{2E}\   
\ln\Bigg\{1+\exp\left[\frac{2\pi}{\hbar\omega}\, 
\left(E - V_{\rm B}\right) \right] \Bigg\}.
\label{sig-Wong}
\end{equation}
For future purposes, we write the above equation in the form,
\begin{equation}
\sigma_{\rm W} = \sigma_0\ F_0(x),
\label{sig-Wong-1}
\end{equation}
where
\begin{equation}
\sigma_0 = \frac{\hbar\omega\,R_{\rm B}^2}{2E}
\label{sig0}
\end{equation}
is a characteristic (energy-dependent) strength of the cross-section, and 
\begin{equation}
F_0 (x)=  \ln\Big[1+e^{2\pi x} \Big]
\label{WFF}
\end{equation}
is the universal fusion function (UFF)~\cite{CGL09a,CGL09b}, which is expressed 
in terms of the dimensionless energy variable
\begin{equation}
x = \frac{E - V_{\rm B}}{\hbar\omega}.
\label{x}
\end{equation}

%%%%%%%%%%%%%%%%%%%%%%%%%%%%%

\subsubsection{The classical limit of the Wong formula}

%%%%%%%%%%%%%%%%%%%%%%%%%%%%

For $2\pi x \gg 1$, one can approximate: $1+\exp (2\pi x) \simeq \exp (2\pi x)$ 
and one gets the classical limit of the universal fusion function
\begin{equation}
 F_0^{\rm cl} (x) = 2\pi x.
\label{sig-Wong1}
\end{equation}
\begin{figure}%[h!]
\begin{center}
\includegraphics*[width=8 cm]{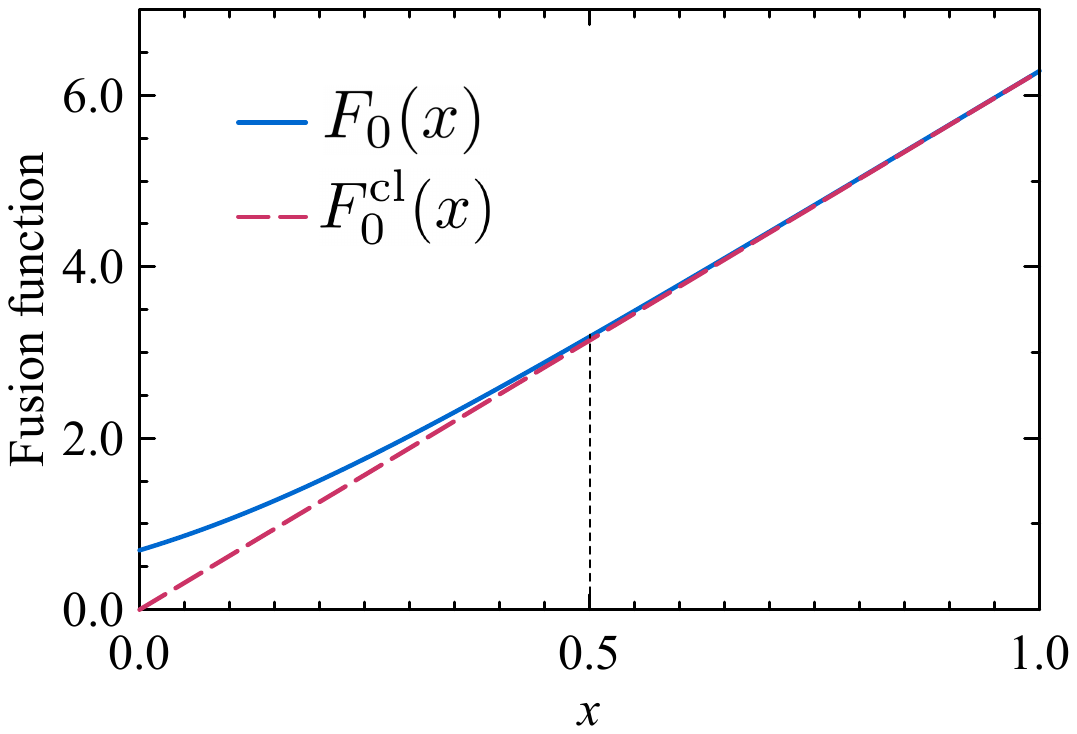}
\end{center}
\caption{The Wong fusion function and its asymptotic limit.}
\label{FF-class}
\end{figure}
The convergence of $F_0 (x)$ to its classical limit is illustrated in 
Fig.~\ref{FF-class}. 
Comparing the two curves, one concludes that the universal fusion function 
can be safely approximated by Eq.~(\ref{sig-Wong1}) for $x \gtrsim 0.5$. 
Since typical values of $\hbar\omega$ are between 2 and 4~MeV, the 
classical cross section of Eq.~(\ref{sigclass}) is very close 
to $\sigma_{\rm W}$, starting at $\sim 1.5$ MeV above $V_{\rm B}$. 
Then, at energies above this limit, we can insert the classical limit of 
the Wong fusion function into Eq.~(\ref{WFF}) and get the classical 
fusion cross-section of Eq.~(\ref{sigclass}), namely
\begin{equation*}
\sigma^{\rm cl}_{\rm W}(E) =  \pi R_{\rm B}^2\ \left( 1 -\frac{V_{\rm B}}{E} 
\right),\qquad {\rm for}\ E\ge V_{\rm B}.
 \end{equation*}
%

%%%%%%%%%%%%%%%%%%%%%%%%%%%%%

\subsubsection{Validity of the Wong formula}

%%%%%%%%%%%%%%%%%%%%%%%%%%%%%

\begin{figure*}%[h!]
\begin{center}
\includegraphics*[width=14 cm]{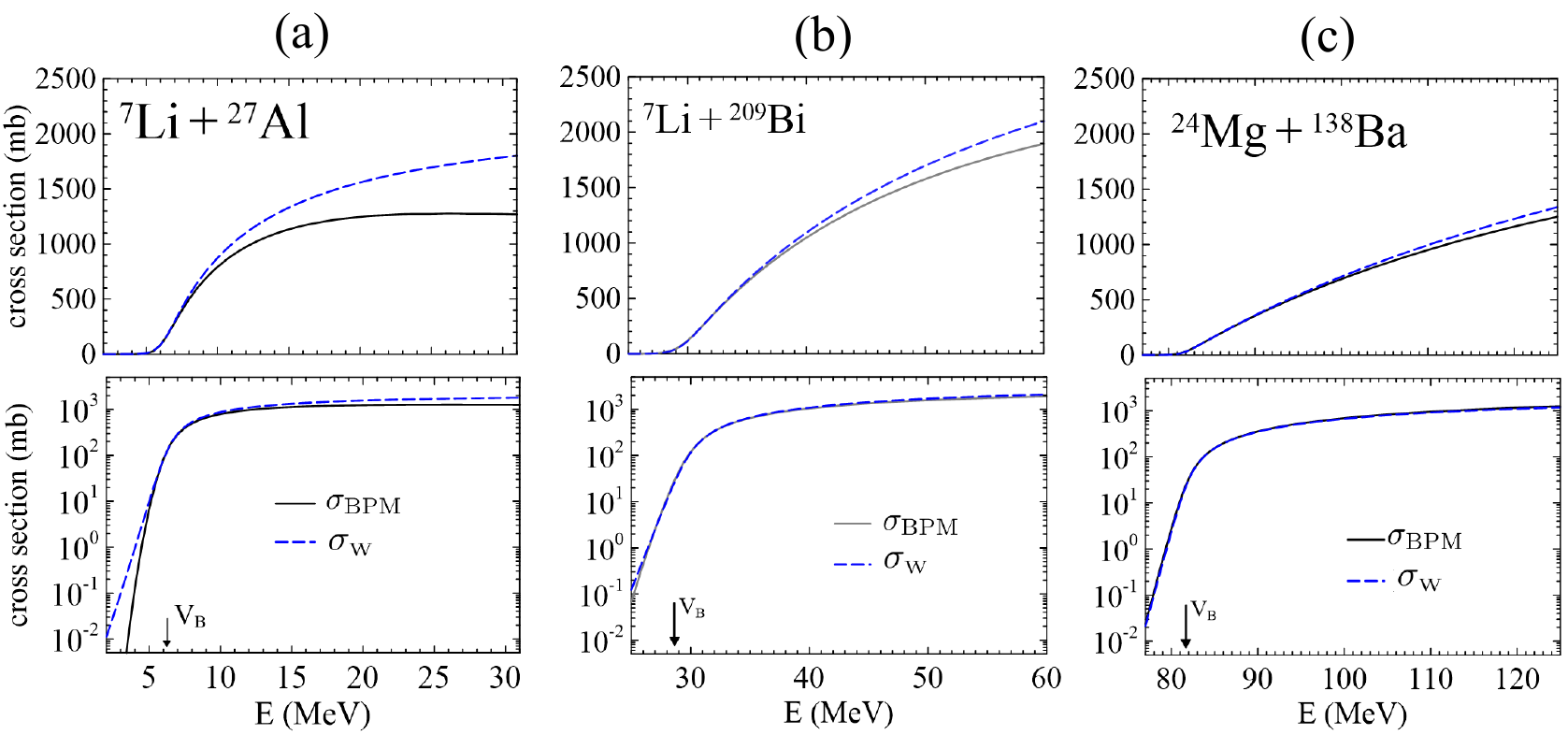}
\end{center}
\caption{The Wong fusion cross section in comparison with the cross sections predicted
by the BPM. See the text for details. }
\label{Wong-vs-QM}
\end{figure*}
The Wong formula is a very nice analytical expression, but its validity is limited. 
It is a very good approximation to the QM fusion cross-section in 
collisions of heavy systems ($Z_{\rm P} Z_{\rm T} > 500$) at near-barrier 
energies~\cite{CGL09b}. 
However, it is not appropriate for light systems 
or at collision energies well below or well above the barrier. \\

The accuracy of the Wong formula is illustrated in Fig.~\ref{Wong-vs-QM}, 
which shows comparisons between $\sigma_{\rm W}$ and $\sigma_{\rm BPM}$ 
for the $^7$Li\,+\, $^{27}$Al ($Z_{\rm P} Z_{\rm T} = 39$),  $^7$Li\,+\,$^{209}$Bi 
($Z_{\rm P} Z_{\rm T} = 249$), 
and $^{24}$Mg\,+\,$^{138}$Ba ($Z_{\rm P} Z_{\rm T} = 672$) systems. 
As in the previous figures, the results are shown at collision energies 
ranging from 4 MeV below the Coulomb barrier to 30 MeV above it. 
Comparing the cross sections for the $^7$Li\,+\,$^{27}$Al system at sub-barrier 
energies, one concludes that the Wong formula overestimates 
$\sigma_{\rm BPM}$ drastically. 
At $E=2$ MeV ($\sim 4$~MeV below $V_{\rm B}$), the Wong cross-section is 
wrong by more than one order of magnitude. 
For the two heavier systems, the Wong formula is quite close to 
$\sigma_{\rm BPM}$  in this energy region.   
 At above-barrier energies, $\sigma_{\rm W}$ is systematically larger  than  
 $\sigma_{\rm BPM}$, mainly in the case of the $^7$Li + $^{27}$Al.  At 30 MeV 
 above $V_{\rm B}$, the Wong cross section for this system overestimates $\sigma_{\rm BPM}$ by
 $\sim 45$\%. The situation is better for the other two heavier systems. For the 
 $^7$Li + $^{209}$Bi and $^{24}$Mg\,+\,$^{138}$Ba systems at the same energy above
$V_{\rm B}$, $\sigma_{\rm W}$ exceeds $\sigma_{\rm BPM}$ by 10\% and $\sim 5\%$, respectively. 
The exceedingly large values of the Wong cross section above the Coulomb barrier  
 can be traced back to neglecting the angular momentum 
 dependence of the barrier parameters. 
 A detailed discussion of the failure of the Wong formula is presented below.\\
\begin{figure}%[h!]
\begin{center}
\includegraphics[width=5 cm]{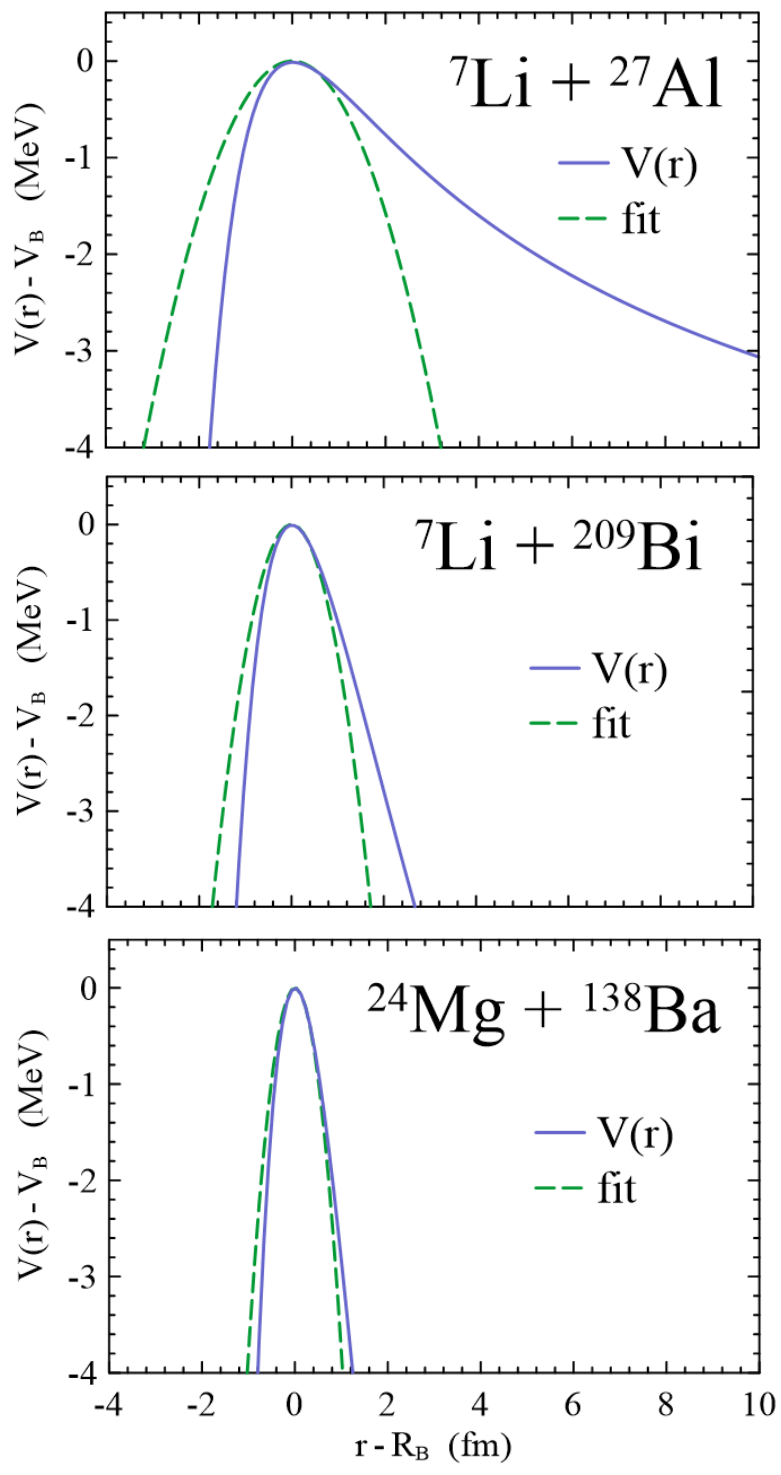}
\end{center}
\caption{The Coulomb barriers for the $^7$Li + $^{27}$Al, $^7$Bi + $^{209}$Al, 
and $^{24}$Mg + $^{138}$Ba systems, in comparison to the corresponding 
parabolic fits.}
\label{par-fits}
\end{figure}

At sub-barrier energies, the cross-section results exclusively from tunneling 
effects. 
Thus, it is susceptible to the shape of the potential barrier. 
Fig.~\ref{par-fits} shows the Coulomb barriers and the parabolic fits for the three 
systems of Fig.~\ref{Wong-vs-QM}. 
The potential axes are truncated at the lowest collision energies of our
calculations, namely $V(r) = V_{\rm B} - 4$~MeV. 
The comparison between the potential barrier of the $^7$Li + $^{27}$Al system 
and the parabolic fit sheds light on the abnormally large values of the 
Wong cross-section at sub-barrier energies. 
The parabolic barrier is much thinner than the actual Coulomb barrier. 
Thus, the transmission coefficient for the parabola is unrealistically large. 
This leads to a considerable enhancement of $\sigma_{\rm W}$ in comparison to 
$\sigma_{\rm BPM}$. 
Conversely, the parabolic fits for the barriers of the two heavier 
systems are quite reasonable. 
Besides, the fitted barrier is thicker on the inner side of the barrier but 
thinner on the outer side. 
Then, there is some compensation in calculating the transmission factors, and
the Wong formula reproduces the QM cross-sections well.\\

Now, we consider the Wong formula above the Coulomb barrier. In this case, the differences between $\sigma_{\rm W}$ and $\sigma_{\rm BPM}$ 
arise from the neglect of the angular momentum dependences of $R_\lambda$ and $\hbar\omega_\lambda$. However, we have shown that the Wong cross-section 
reduces to the classical cross-section just above the Coulomb barrier, and this cross section does not depend on the barrier curvature. Then, we 
concentrate on the angular momentum dependence of $R_\lambda$.\\

\begin{table}%[h!]
\centering
\caption{Variation of the barrier radii as the angular momentum varies from 0 to its grazing value for the systems discussed in the text. Column two gives the maximal energy considered
in our calculations. For each system, it corresponds to the Coulomb barrier plus 30 MeV. The remaining columns are explained in the text. The grazing angular momenta are given in 
$\hbar$ units; the barrier radii are given in fm.}
\vspace{0.5cm}
\begin{tabular}{lcccccc}
\hline 
System                               \ \ \ &  $E_{\rm max}$  \ \ \ \ &  $\lambda_{\rm g}$ \ \ \ \  &   $R_{\rm g}$ \ \ \   &  $R_{\rm B}$  \ \ \ & $R^2_{\rm g}/R^2_{\rm B}$ \\ 
\hline
$^7$Li\ +\ $^{27}$Al             \ \ \  &    36                   \ \ \ \ &  20                      \ \ \ \  & 6.2                 \ \ \   &  8.4                   \ \ \ &                            0.53             \\
%\hline  
$^7$Li\ +\ $^{209}$Bi           \ \ \  &    60                   \ \ \ \ &  34                     \ \ \ \   & 10.4                \ \ \  & 11.4                  \ \ \ &                            0.83            \\
%\hline  
$^{24}$Mg\,+$^{138}$Ba   \ \ \  &   112                 \ \ \ \  &  59                       \ \ \ \ &  10.6               \ \ \  &  11.1                  \ \ \   &                            0.91              \\
%\hline  
%
\hline
\end{tabular}
\label{barrier parameters}
\end{table}
Table \ref{barrier parameters} shows the explicit value of the highest energy considered in our calculations for each system, namely 
$E_{\rm max} = V_{\rm B} + 30$ MeV. 
The next two columns show the corresponding values of the grazing angular momentum and the barrier radius associated with it. They 
are represented by $\lambda_{\rm g}$ and $R_{\rm g}$, respectively.
The table also shows the s-wave barrier radius and the ratio $R^2_{\rm g}/R^2_{\rm B}$. This ratio estimates the inaccuracy 
of Wong and the classical expressions in the worst scenario of the present calculations.\\

In the $^7$Li + $^{27}$Al collision at $E=36$ MeV, the angular momentum dependence of $R_\lambda$ is significant. If one uses
$R_{\rm g}^2$ instead of $R_{\rm B}^2$ in the Wong or the classical expression, as proposed by Rowley and Hagino~\cite{RoH15}, the cross section 
is reduced by a factor of $\sim 2$. The situation is better for the $^7$Li\ +\ $^{209}$Bi and mainly the $^{24}$Mg\,+$^{138}$Ba systems, where the 
reduction factors would be considerably closer to one (0.83 and 0.91, respectively).

%%%%%%%%%%%%%%%%%%%%%%%%%%%%%%%%%%%%%%%%%%%%%%%%%%%%%%

\section{Improved versions of the Wong and the classical cross sections}
\label{improv-Wong}

%%%%%%%%%%%%%%%%%%%%%%%%%%%%%%%%%%%%%%%%%%%%%%%%%%%%%%
The previous sub-section showed that the Wong formula works poorly 
for light systems at energies well below or above the Coulomb 
barrier. The problem can be fixed by introducing effective barrier parameters, 
$\overline{R}$ and $\hbar\bar{\omega}$, in the modified Wong formula,\\
\begin{equation}
\overline{\sigma}_{\rm W} =\frac{\hbar\bar{\omega}\,
\overline{R}^{\,2}}{2E}\   \ln\Bigg\{
1+\exp\left[\frac{2\pi}{\hbar\bar{\omega}}\, 
\left(E - V_{\rm B}\right) \right] \Bigg\}.
\label{Wong-new}
\end{equation}

Since the behavior of the cross section at sub-barrier energies is determined 
by the argument of the exponential,
$2\pi\,\left(E-V_{\rm B}\right)/\hbar \bar{\omega}$, we keep the original values 
of the barrier parameters in the slowly varying multiplicative factor. 
That is, we approximate
\begin{equation}
\frac{\hbar\bar{\omega}\,\overline{R}^2}{2E} \simeq \frac{\hbar\omega\,R_{\rm B}^2}{2E}.
\label{Wong-improved}
\end{equation}
Then, we get the effective barrier curvature parameter, $\hbar\bar{w}$, by imposing that the 
modified Wong cross section of Eq.~(\ref{Wong-new}) be equal to the fusion cross section of 
the barrier penetration model. We find,\\
\begin{equation}
\hbar\bar{\omega}=  2\pi \left( E-V_{\rm B}\right)/\;
{\rm ln}\left[
\exp \left(
\frac{2 E\, \sigma_{\rm BPM}}{\hbar\omega\, R_{\rm B}^2} -1
\right)
\right]. 
\label{hweff-subbar}
\end{equation}

At above-barrier energies, the main contributions to the fusion cross-section 
come from angular momenta in the vicinity of $\lambda_{\rm g}$. 
This led Rowley and Hagino~\cite{RoH15} to propose an improved version 
of the Wong formula. 
It consists of replacing $R_{\rm B}$ and $\hbar\omega$ with barrier parameters 
for $\lambda_{\rm g}$. 
The Wong formula then becomes
\begin{equation}
\sigma_{\rm W}^{\rm g} =\frac{\hbar\omega_{\rm g}\,R_{\rm g}^2}{2E}\   
\ln\Bigg\{1+\exp\left[\frac{2\pi}{\hbar\omega_{\rm g}}\, 
\left(E - V_{\rm B}\right) \right] \Bigg\},
\label{Wong-RH}
\end{equation}
where we used the short-hand notations: 
$\hbar\omega_{\rm g} \equiv \hbar\omega_{l_{\rm g}}$.\\
\begin{figure*}%[h!]
\begin{center}
\includegraphics[width=18 cm]{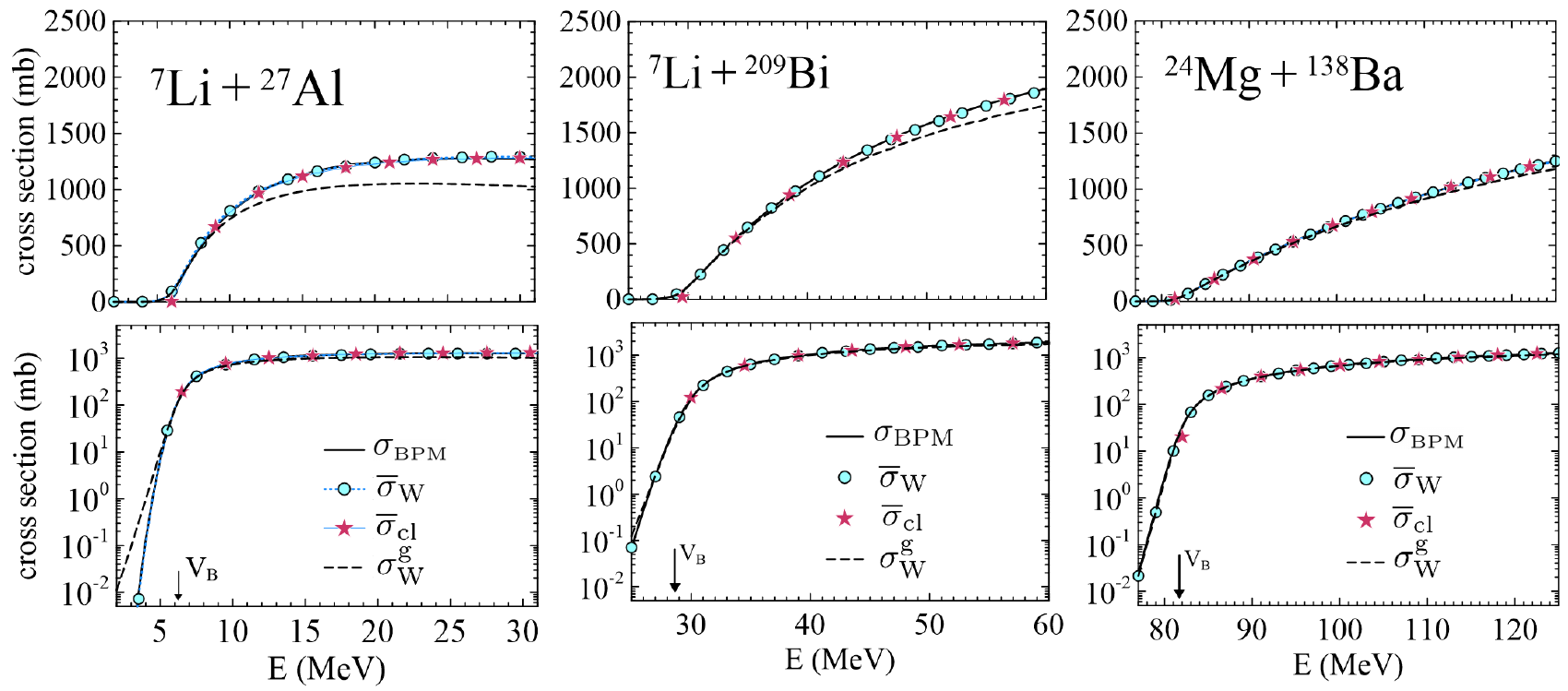}
\end{center}
\caption{The BPM fusion cross sections for the $^7$Li + $^{27}$Al, 
$^7$Li + $^{209}$Bi, and $^{24}$Mg + $^{138}$Ba systems, in comparison 
to the ones obtained by the improved versions of the Wong formula, 
$\sigma_{\rm W}^{\rm g}$ and $\overline{\sigma}_{\rm W}$. 
The improved classical cross-sections, 
$\overline{\sigma}_{\rm cl}$, are also shown.}
\label{Wong improved-vs-QM}
\end{figure*}

As can be seen in Table \ref{barrier parameters}, the barrier radius 
decreases as $\lambda$ increases, and replacing $R_{\rm B}^2$ with 
$R_{\rm g}^2$ in the Wong formula reduces the cross-section at 
above-barrier energies. 
Since the Wong formula overestimates $\sigma_{\rm BPM}$ 
(see Fig.~\ref{Wong-vs-QM}), this modification is expected to improve 
the agreement between the two cross sections. 
However, it might overestimate the weight of the grazing angular momentum in the
partial-wave series. 
This possibility is avoided in the improved Wong cross section proposed below.\\

Since the system Hamiltonian depends on $\lambda$ quadratically (through
the centrifugal term of the potential), we assume that the angular momentum
dependence of the barrier radius is also quadratic, at least around $\lambda=0$.  
Then, we make a series expansion of $R_\lambda$ and keep only the term of 
lowest order. 
We get
\begin{equation}
R_\lambda \simeq R_{\rm B} - \gamma\,\lambda^2,
\label{Rl-expansion}
\end{equation}
where $\gamma$ is a system-dependent parameter. \\

Then, we define the effective barrier radius in a collision with energy $E$ 
and grazing angular momentum $\lambda_{\rm g}$ as the weighted average
\begin{equation}
\overline{R} =  \frac{1}{N}\, \int_{0}^{\lambda_{\rm g}}  2 \lambda\,d\lambda\ 
\left[ R_{\rm B} - \gamma\,\lambda^2\right],
\label{average-0}
\end{equation}
where $N$ is the norm
\begin{equation}
N =  \int_{0}^{\lambda_{\rm g}}  2 \lambda\,d\lambda.
\label{norm}
\end{equation}
At above-barrier energies,  the integrations involve large values of $\lambda$. 
Then, it is a reasonable approximation to use $\lambda=0$ (instead of 
$\lambda =1/2$) as the lower limit of the integrations in 
Eqs.~(\ref{average-0}) and (\ref{norm}). 
Following this procedure, we get
\begin{equation}
\overline{R} = R_{\rm B} - \gamma\,\frac{\lambda_{\rm g}^2}{2}.
\label{Rbar-exp}
\end{equation}
Comparing the above expression with Eq.~(\ref{Rl-expansion}), one concludes
that $\overline{R}$ corresponds to the  barrier radius of the $\lambda$-dependent 
potential of Eq.~(\ref{Vlambda}) at the effective angular momentum 
\begin{equation}
\overline{\lambda} = \frac{\lambda_{\rm g}}{\sqrt{2}} .
\label{lambda-bar}
\end{equation}
Thus, we can write
\begin{equation}
\overline{R} = R_{\overline{\lambda}}.
\label{Rbar0}
\end{equation}
Note that $\overline{R}$ is fully determined by Eqs.~(\ref{Rbar0}) 
and (\ref{lambda-bar}). 
Therefore, one does not need the explicit value of the coefficient 
$\gamma$ in the expansion of Eq.~(\ref{Rl-expansion}).\\

The curvature parameter can be modified in the same way, namely
\begin{equation}
 \hbar\bar{\omega} = \hbar\omega_{\,\overline \lambda}.
\label{hwbar}
\end{equation}
The improved Wong cross section at above-barrier energies is then given 
by Eq.~(\ref{Wong-new}), with the $\overline{R}$ and
$\hbar\overline{\omega}$ parameters of the above equations. \\

An improved version of the classical cross-section of 
Eq.~(\ref{sigclass}) can be derived by the same procedure. 
Replacing $R_{\rm B}$ with $\overline{R}$, one gets \\
\begin{eqnarray}
\overline{\sigma}_{\rm cl} &=&  \pi \overline{R}^2\ 
\left( 1 -\frac{V_{\rm B}}{E} \right),\qquad {\rm for}\ E\ge V_{\rm B};\nonumber\\
 &=&  0 \qquad\qquad \qquad\qquad\ \ \ \ {\rm for}\ E < V_{\rm B}.
\label{sigclass-new}
\end{eqnarray}
It is convenient to introduce the dimensionless energy variable,
\begin{equation}
y = 1\ -\ \frac{V_{\rm B}}{E}.
\label{y}
\end{equation}
Then, the standard classical cross section takes the form
\begin{equation}
\sigma_{\rm cl} = \pi\,R_{\rm B}^2 \ \, y,
\label{sigclass-y}
\end{equation}
and its improved version can be written as
\begin{equation}
\overline{\sigma}_{\rm cl} = f_{\rm R}(y) \times \sigma_{\rm cl}.
\label{sigclass-new1}
\end{equation}
Above, $f_{\rm R}(y)$ is the correction factor
\begin{equation}
f_{\rm R}(y) = \left[ \frac{\overline{R}(y)}
{R_{\rm B} }\right]^2,
\label{fRy}
\end{equation}
which is always less than one.
To stress the energy-dependence (or $y$-dependence) of $\overline{R}$, 
we used the notation $\overline{R}(y)$. 
One may notice that the $\overline{R}(y)$ value may be obtained directly 
from the $f_{\rm R}(y)$ function. 
The next subsection will present an empirically obtained version of this function.\\
Figure ~\ref{Wong improved-vs-QM} shows the approximate cross sections 
$\sigma_{\rm W}^{\rm g}$ (Eq.~(\ref{Wong-RH})), 
$\overline{\sigma}_{\rm W}$ (Eq.~(\ref{Wong-new})) and 
$\overline{\sigma}_{\rm cl}$ (Eq.~(\ref{sigclass-new1})), for the
$^7$Li + $^{27}$Al, $^7$Li + $^{209}$Bi, and $^{24}$Mg + $^{138}$Ba systems. 
They are represented by dashed lines, blue circles, and red stars, respectively. 
The solid lines correspond to the BPM cross sections. 
First, one notices that $\sigma_{\rm W}^{\rm g}$ (dashed lines) 
actually underestimates $\sigma_{\rm BPM}$ at above-barrier energies. 
The difference between the two cross-sections is particularly large in 
the case of the light $^7$Li + $^{27}$Al system. 
The difference is smaller for the $^7$Li + $^{209}$Bi system, and the agreement 
is very good for $^{24}$Mg + $^{138}$Ba. 
On the other hand, the improved Wong fusion cross-sections 
($\overline{\sigma}_{\rm W}$) for the three systems are in excellent 
agreement with the corresponding $\sigma_{\rm BPM}$ below and above 
the Coulomb barrier. 
One also observes that, at above-barrier energies, the improved classical 
cross sections, $\overline{\sigma}_{\rm cl}$, reproduce $\sigma_{\rm BPM}$ 
equally well, except in a very small energy interval just above $V_{\rm B}$.

%%%%%%%%%%%%%%%%%%%%%%%%%%%%%%%%%%%%%%%%%%%%%%%%%%%

\subsection{An approximate expression for $\overline{R}$}

%%%%%%%%%%%%%%%%%%%%%%%%%%%%%%%%%%%%%%%%%%%%%%%%%%%

The barrier parameters of the real potential, $V_{\rm B}$, $R_{\rm B}$, 
and $\hbar\omega$, can be obtained from available computer codes. 
However, the effective barrier radius, $\overline{R}$, is more complicated
to obtain, as it is necessary to determine the function $f_{\rm R}(y)$, of 
Eq.~(\ref{fRy}). \\

\begin{figure}[t!]
\begin{center}
\includegraphics[width=8.2cm]{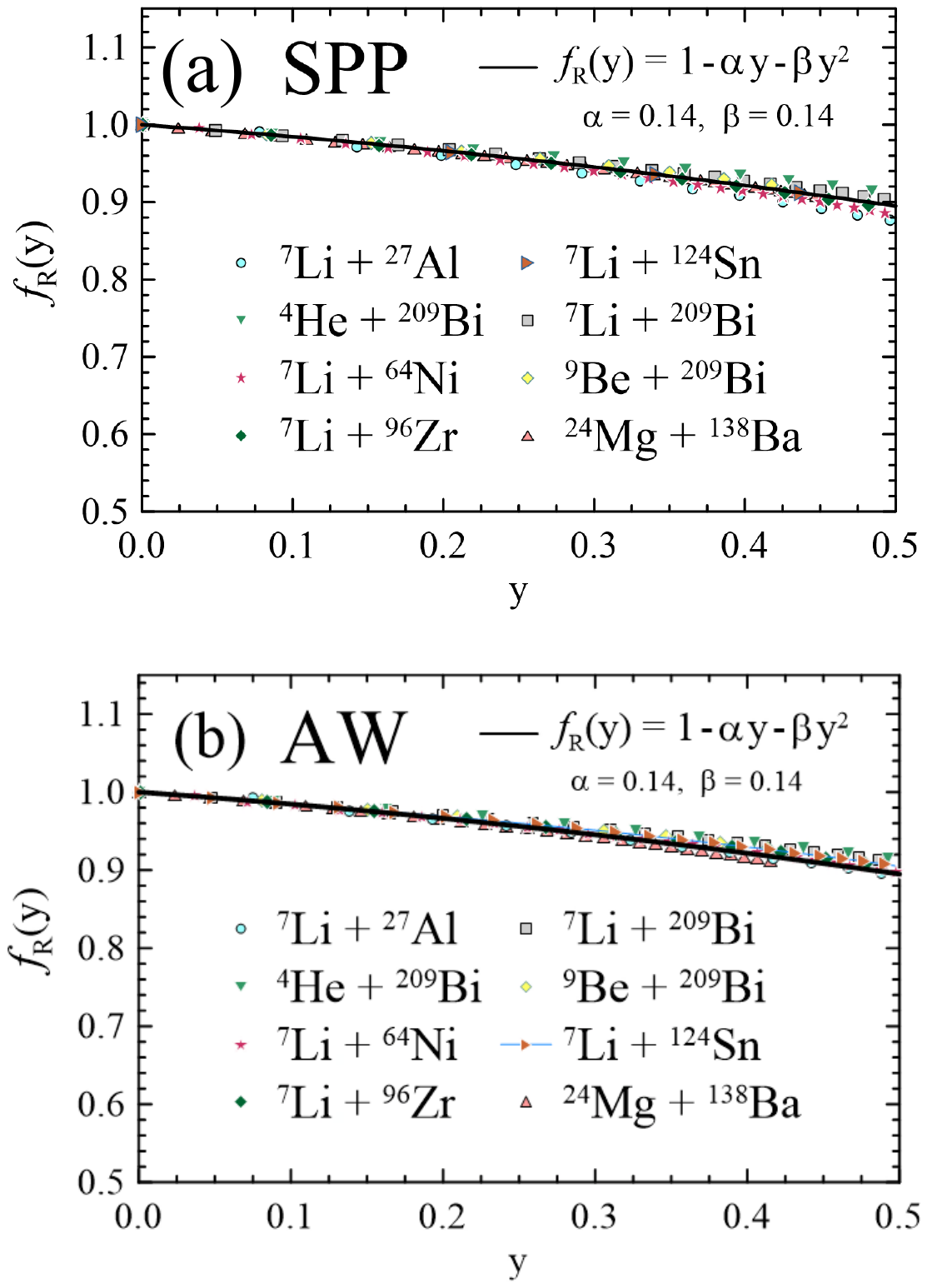}
\end{center}
\caption{The function $f_{\rm R}(y)$ of Eq.~(\ref{fRy}), for the $^7$Li + 
$^{27}$Al, $^7$Li + $^{209}$Bi, and $^{24}$Mg + $^{138}$Ba systems, plotted
versus the dimensionless energy variable, $y$}.
\label{fRw-fig}
\end{figure}

We carried out a systematic study of this function, considering several 
systems over a broad mass range, and two commonly used nuclear interactions: 
the SPP~\cite{CPH97,CCG02} and the AW~\cite{AkW81,BrW04} potentials. 
In each case, we evaluated $f_{\rm R}(y)$ for collision energies ranging 
from $E = V_{\rm B}$ to $E = 2\times V_{\rm B}$. 
The results for the SPP and for the AW interactions are shown in 
Figs.~\ref{fRw-fig}(a) and \ref{fRw-fig}(b), respectively. 
The points correspond to the effective barrier radius numerically 
calculated at a mesh of collision energies. 
The solid line is the empirical function 
\begin{equation}
f(y)\sim f_{\rm app}(y) = 1\,-\,0.14 y\,-\,0.14\,y^2,
\label{ffit}
\end{equation}
which gives the best fit to the data. 
Inspecting Figs.~\ref{fRw-fig}(a) and \ref{fRw-fig}(b), one arrives 
at two conclusions. 
The first is that $f_{\rm R}(y)$ has a very weak dependence on the system. 
For all systems, $f_{\rm R}(y)$ is given by Eq.~(\ref{ffit}) as an excellent 
approximation. 
The second conclusion is that $f_{\rm R}(y)$ is not very sensitive
to the choice of nuclear potential. 
For the SPP (Fig.~\ref{fRw-fig}(a)) and the AW (Fig.~\ref{fRw-fig}(b)) 
potentials, $f_{\rm R}(y)$ is reproduced by $f_{\rm app}(y)$ equally well. 
Thus, the influence of the real potential is exerted exclusively
through the value of the radius parameter, $R_{\rm B}$. \\

%

%%%%%%%%%%%%%%%%%%%%%%%%%%%%%%%%%%%%%%%%%%%%%%%%%%%%%%%%%%%%%%%%%%%%%%%%

\subsection{Reduction of fusion data and universal functions}
\label{sec-Univfun}

%%%%%%%%%%%%%%%%%%%%%%%%%%%%%%%%%%%%%%%%%%%%%%%%%%%%%%%%%%%%%%%%%%%%%%%%

A frequently used reduction procedure is the fusion function (FF) method, 
which is based on the Wong formula. 
The collision energy, $E$, and the fusion cross section, $\sigma_{\rm F}$, 
are transformed into the dimensionless quantities
\begin{equation}
E\, \rightarrow\, x=\frac{E-V_{\rm B}}{\hbar\omega},\qquad 
\sigma_{\rm F} \, \rightarrow\, F(x) = \frac{\sigma_{\rm F}}{\sigma_0},
\label{FF-reduc}
\end{equation}
where $\sigma_0$ is the characteristic cross section of Eq.~(\ref{sig0}). \\

The improved Wong cross-section of Eq.~(\ref{Wong-new}) leads to an improved 
fusion function (IFF) method, implemented by the transformations
\begin{equation}
E\, \rightarrow\, \overline{x} = \frac{E-V_{\rm B}}{\hbar\bar{\omega}},
\qquad \sigma_{\rm F} \, \rightarrow\, 
\overline{F}(x) = \frac{\sigma_{\rm F}}{\overline{\sigma}_0}.
\label{FF-reduc-imp}
\end{equation}
Above, $\hbar\bar{\omega}$ is the effective barrier curvature parameter of 
Eq.~(\ref{hweff-subbar}), and $\overline {\sigma}_0$ is the characteristic
cross section of the improved Wong formula,
\begin{equation}
\overline{\sigma}_0 = \frac{\hbar\overline{\omega}\,\overline{R}^2}{2E} .
\label{sigbar0}
\end{equation}
\begin{figure}%[h!]
\begin{center}
\includegraphics[width=8.2 cm]{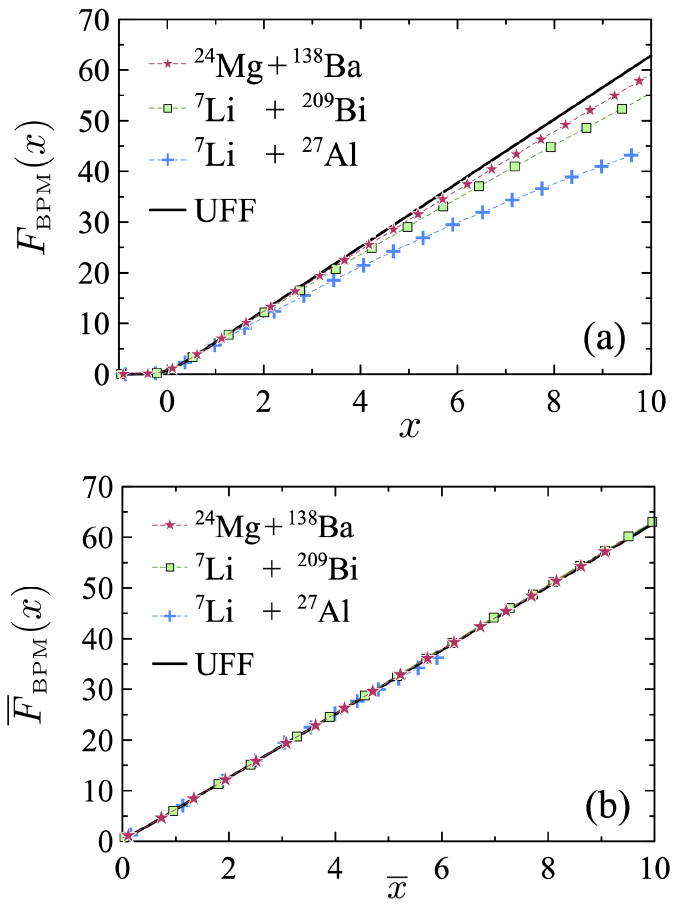}
\end{center}
\caption{Reduced BPM cross sections for the $^7$Li + $^{27}$Al, 
$^7$Li + $^{209}$Bi and $^{24}$Mg + $^{138}$Ba systems, reduced by the two fusion function procedures discussed in the text.}
\label{FF and FFimprov}
\end{figure}
As a simple test, the reduction procedures can be applied to the BPM 
fusion cross section~\cite{CGL09b,CMG15}. 
The procedure is successful if reduced cross sections for systems in 
different mass ranges are very similar. 
Further, if the procedure leads to a universal function, the reduced 
cross-sections should be very close to it. 
We applied this test to the two versions of the fusion function method,
Eqs.~(\ref{FF-reduc}) and (\ref{FF-reduc-imp}). \\

Fig.~\ref{FF and FFimprov}(a) shows the reduced $\sigma_{\rm BPM}$ 
fusion cross sections for the $^7$Li + $^{27}$Al, 
$^7$Bi + $^{209}$Al, and $^{24}$Mg + $^{138}$Ba systems. 
They are denoted by $F_{\rm BPM}(x)$. 
The reduction was carried out through the standard fusion function method 
of Eq.~(\ref{FF-reduc}). 
For comparison, the UFF is also shown (black solid line). 
One notices that the fusion functions exhibit a significant system dependence.  
At the highest energies, corresponding to $x\sim 10$ ($\sim 25$~MeV above 
$V_{\rm B}$), the fusion function for the $^7$Li + $^{27}$Al system is 
$\sim 30\,\%$ lower than the UFF. 
The fusion functions for the two heavier systems remain below the UFF, 
but the difference is much smaller. 
Despite this system dependence, the fusion method has been widely used in 
comparing fusion data of weakly bound 
systems~\cite{CGL09b,GLC09,GCL11,WZG14,GAA21,YSZ21,BSA22}.
In practical studies of nuclear structure effects based on this 
reduction method, the fusion functions are renormalized to avoid system 
dependencies arising from the inaccuracy of the Wong formula~\cite{CGL09b}.\\

Next, we apply the same test to the improved fusion function method of 
Eq.~(\ref{FF-reduc-imp}). 
The reduced cross sections, shown in Fig.~\ref{FF and FFimprov}(b), 
are denoted by $\overline{F}_{\rm BPM}(\overline{x})$. 
Now, the situation is entirely different. 
The system dependence of the previous figure is fully eliminated. 
The fusion functions for different systems can hardly be distinguished, 
and they agree very well with the UFF. \\
%We conclude that the improved fusion 
%function method is expected to be a very useful tool for comparative 
%studies of fusion data.\\

The fact that the improved classical cross section of Eq.~(\ref{sigclass-new}) 
is very close to $\sigma_{\rm BPM}$ leads to another universal
function, which we denote by $G_0(y)$. 
It is obtained by the transformations
\begin{equation}
E\,\rightarrow\, y = 1 - \frac{V_{\rm B}}{E},\qquad 
\overline{\sigma}_{\rm cl}\,\rightarrow\, 
G_0 = \frac{ \overline{\sigma}_{\rm cl}} {\pi \overline{R}^2}.
\label{y-G0}
\end{equation}
In this way, one gets the {\it classical fusion line} (CFL)
\begin{equation}
G_0 (y) = y.
\label{G0(y)}
\end{equation}
%
%That is, a linear function passing through the origin, with angular 
%coefficient $\alpha = 1$.\\

The above transformation suggests a new reduction procedure to analyze 
fusion data at energies above the Coulomb barrier. 
The energy is transformed into the dimensionless energy variable, $y$, 
as in Eq.~(\ref{y-G0}), and the fusion cross section, $\sigma_{\rm F}$, 
is transformed into the {\it classical fusion function} (CFF), \\
\begin{equation}
\overline{G}(y) = \frac{\sigma_{\rm F}}{ \pi \overline{R}^2}
= \frac{\sigma_{\rm F}}{ \pi R_{\rm B}^2 \,f_{\rm R}(y)}.
\label{G(y)}
\end{equation}
Above, $f_{\rm R}(y)$ is the ratio $\overline{R}^2/R_{\rm B}^2$, 
introduced in Eq.~(\ref{fRy}). \\

\begin{figure}[t!]
\begin{center}
\includegraphics[width=8.2cm]{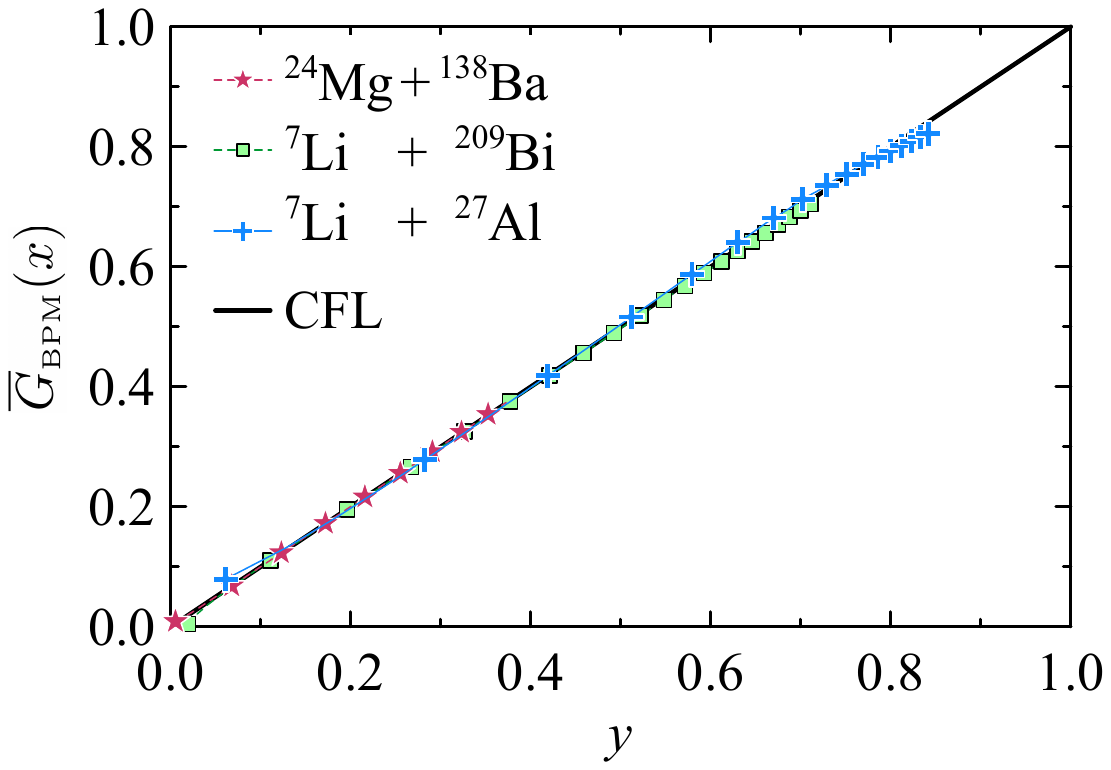}
\end{center}
\caption{Reduced BPM cross sections for the same systems as the previous figure,
but the CFF method now carries out the reductions.}
\label{RCF}
\end{figure}
We submitted the classical fusion function (CFF) reduction method of 
Eqs.~(\ref{y-G0}) and (\ref{G(y)}) to the same test applied to the fusion function method. 
Since the empirical function $f_{\rm app}(y)$ is quite close to
$f_{\rm R}(y)$, we used the approximate expression 
\begin{equation}
\overline{G}^{\,\rm exp}(y) = \frac{\sigma_{\rm F}^{\,\rm exp}}{\pi\,f_{\rm app}(y)\,R_{\rm B}^2}.
\label{Gexp-app}
\end{equation}

The reduced  $\sigma_{\rm BPM}$ cross sections, denoted by 
$\overline{G}_{\rm BPM}(y)$, for the $^7$Li + $^{27}$Al, $^7$Bi + $^{209}$Bi, 
and $^{24}$Mg + $^{138}$Ba systems are shown in Fig.~\ref{RCF}.
The black solid line represents the classical fusion line of Eq.~(\ref{G0(y)}). 
We observe that the reduced cross-sections for the three systems are extremely 
close and agree very well with the classical fusion line. \\

An interesting point in Fig.~\ref{RCF} is that the results for heavier systems 
are concentrated at the lower region of the CFL, whereas those for lighter systems reach the high end of this line. 
The reason is that our calculations were performed at energies up to  
$\sim V_{\rm B}+30$~MeV. 
In this way, the largest $y$-value in the calculations is
\[
y_{\rm max} = 1\ - \ \frac{V_{\rm B}}{V_{\rm B} + 30\,{\rm MeV}} . 
\]
Then, for the lightest systems, $^7$Li + $^{27}$Al, the barrier 
($V_{\rm B} = 6.1$~MeV) is much lower than 30 MeV and $y_{\rm max}$ 
is close to one.
For the heaviest system, $^{24}$Mg + $^{138}$Ba, the barrier 
($V_{\rm B} = 81,6$~MeV) is much larger than 30 MeV, so that 
$y_{\rm max}$ is much lower. \\

Note that, above the Coulomb barrier, the fusion cross section does not 
depend on the barrier curvature. 
This can be seen explicitly in the expression for $\overline{\sigma}_{\rm cl}$but needs to be clarified in the Wong formula. 
However, it becomes transparent if one uses the classical limit of $F_0(x)$ 
(Eq.~(\ref{sig-Wong1})). 
Then, the dependence on $\hbar\omega$ cancels when $F_0(x)$ is multiplied 
by $\sigma_0$.\\

%%%%%%%%%%%%%%%%%%%%%%%%%%%%%%%%%%%%%%%%%%%%%%%%%%%%%%%

\section{Application: reduced cross sections of weakly bound systems}
\label{app}

%%%%%%%%%%%%%%%%%%%%%%%%%%%%%%%%%%%%%%%%%%%%%%%%%%%%%%%

The experimental fusion functions evaluated by the IFF method are 
very similar to the ones reduced by the standard FF method, reported
in Ref.~\cite{CGL09a}. 
The advantage of the former over the latter method is that it does not 
require a renormalization to take care of the inaccuracies of the Wong 
formula. 
However, the results of the two methods are very similar. 
For this reason, the discussion of the present section will be focused 
on the CFF reduction method. \\

We used the CFF method to reduce the fusion data of several tightly and 
weakly bound systems, performing the transformations
\begin{equation}
E\ \longrightarrow\ y=1-\frac{V_{\rm B}}{E},\qquad \  
\overline{G}^{\,\rm exp}(y) = \frac{\sigma_{\,\rm F}^{\rm exp}}
{\pi\,\overline{R}^2}.
\label{Gexp}
\end{equation}
To accomplish this goal, one needs the barrier height of the potential, 
$V_{\rm B}$, and the effective barrier radius, $\overline{R}$.

\subsection{Comparative study of experimental CFFs}

In this section, we apply the CFF reduction method to the fusion data of 
weakly bound systems. \\

Before considering weakly bound systems, we submitted the CFF method 
to a test similar to the one carried out in Sect.~\ref{sec-Univfun}. 
The difference is that instead of applying it to BPM cross sections, 
we reduce experimental fusion cross sections of tightly bound systems 
where nuclear structure properties do not influence the cross section at above-barrier energies. 
In such cases, one expects the reduced fusion data to be very 
close to the classical fusion line of Eq.~(\ref{G0(y)}).\\

\begin{figure}[t!]
\begin{center}
\includegraphics[width=8.2cm]{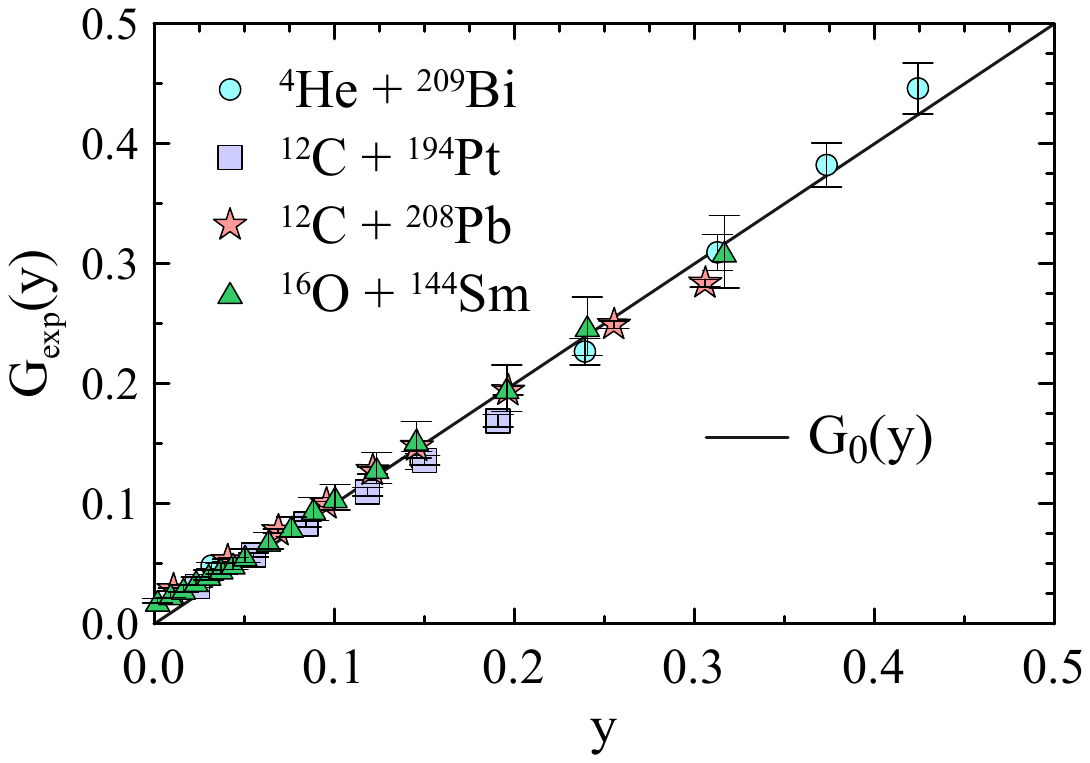}
\end{center}
\caption{Experimental CFF for a few tightly bound systems with negligible 
channel coupling effects.}
\label{CFF-TB}
\end{figure}
Fig.~\ref{CFF-TB} shows the experimental CFF for the 
$^4$He + $^{209}$Bi~\cite{HLK06}, $^{12}$C + $^{194}$Pt~\cite{SKC01}, 
$^{12}$C + $^{208}$Pb~\cite{MHD07}, and $^{16}$O + $^{144}$ Sm~\cite{LDH95} 
systems. 
To obtain the CFFs of the figure, $R_{\rm B}$ was derived using 
the S\~ao Paulo potential with realistic nuclear densities 
and $f_{\rm R}(y)$ by the pocket formula of Eq.~(\ref{ffit}). 
One observes that the reduced data points are, indeed, very close to the 
classical fusion line. \\

Now, we study CFFs associated with the CF of weakly bound projectiles. 
As we will show below, the reduced data of these systems 
can be fitted by a linear function through the origin of the form
\begin{equation}
\overline{G}_{\rm exp}(y) = \alpha\ y.
\label{linear}
\end{equation}
Since the classical fusion line is given by this function with $\alpha = 1$, 
the angular coefficient for a weakly bound projectile
can be interpreted as a CF survival factor. This probability is usually called the {\it suppression factor}. We do not adopt this terminology here, because it is not intuitive: 
a large suppression factor corresponds, actually, to a small suppression, and the other way
around.\\
\begin{figure*}[t!]
\begin{center}
\includegraphics[width=16 cm]{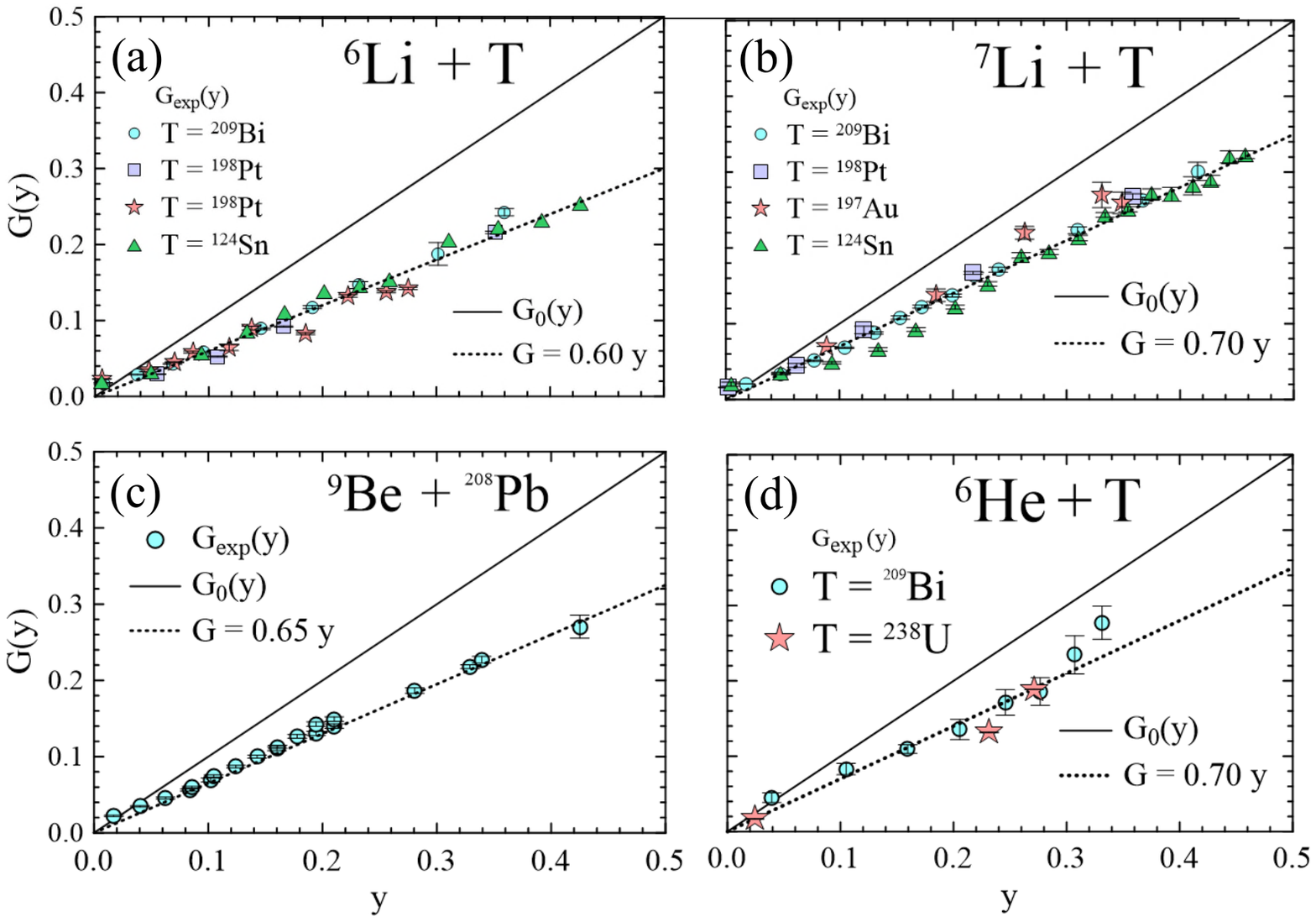}
\end{center}
\caption{Experimental CFF for weakly bound projectiles with different targets: 
(a) collisions of $^6$Li, (b) collisions of  $^7$Li, (c) collisions of $^9$Be, 
(d) collisions of $^6$He. 
See the text for details.}
\label{all-wb}
\end{figure*}
Fig.~\ref{all-wb} shows experimental CFFs for several weakly bound systems. 
In Fig.~\ref{all-wb}[a], the projectile is $^6$Li, which has a breakup 
threshold of 1.47 MeV, and the targets are $^{209}$Bi, $^{198}$Pt, $^{197}$Au, 
$^{124}$Sn, and $^{90}$. 
The data are from Refs.~\cite{DHH02,DGH04} ($^6$Li + $^{209}$Bi), \cite{SNL09} 
($^6$Li + $^{198}$Pt), \cite{PTN14} ($^6$Li + $^{197}$Au), 
\cite{PPS18} ($^6$Li +  $^{124}$Sn), and \cite{KJP12} ($^6$Li + $^{90}$Zr).
Note that we are considering targets over a significant mass range. 
The atomic and mass numbers of the targets are comprised of the 
intervals $\{40,83\}$ and $\{124,209\}$, respectively. 
One finds that the experimental CFF of these systems are very similar. 
They follow closely the linear function of Eq.~(\ref{linear}), with the CF survival factor\\
\begin{equation}
\alpha^{^6{\rm Li}} = 0.60.
\end{equation}

Next, we consider collisions of $^7$Li projectiles, which have a breakup 
threshold of 2.47 MeV. 
Fig.~\ref{all-wb}[b]  shows the CFFs corresponding to the
fusion data for the $^{209}$Bi~\cite{DHH02,DGH04}, $^{198}$Pt~\cite{SND13}, 
$^{197}$Au~\cite{PTN14,nndc}, and $^{124}$Sn~\cite{PPS18} targets. 
Qualitatively, the figure is very similar to the previous one. 
The experimental CFF points for these targets are also distributed along a 
straight line but with a slightly larger CF survival factor\\
\begin{equation}
\alpha^{^7{\rm Li}} = 0.70.
\label{alpha07}
\end{equation}

Comparing Figs.~\ref{all-wb}(a) and \ref{all-wb}(b), one observes 
that the CF survival factor decreases with the breakup threshold of the 
projectile, as expected. 
However, these figures reveal another very interesting behavior of the 
CF suppression: it seems independent of the target charge. 
The results for $^{90}$Zr ($Z_{\rm T}=40$) are essentially the same as 
the ones for the $^{209}$Bi target ($Z_{\rm T}=83$), which has an atomic 
number about twice as large. 
Note that the same conclusion was reached in the study of Kumawat 
{\it et al.}~\cite{KJP12}. 
This finding suggests that the Coulomb breakup does not affect the CF 
cross section. 
Thus, one is led to the conclusion that the observed CF suppression in 
collisions of weakly bound projectiles with heavy targets is due to 
nuclear breakup. 
This conclusion seems to contradict our knowledge of breakup reactions, 
which are strongly influenced by Coulomb breakup 
and Coulomb-nuclear interference~\cite{NuT98,NuT99,OGL13,DCH17}. \\

However, there is a significant difference between the breakup and 
the fusion processes. 
The former gets contributions from hundreds of partial waves, most
corresponding to distant collisions, where the distance of closest approach 
is larger than the range of the nuclear potential. 
At lower partial waves, nuclear couplings are also relevant, and there are 
strong Coulomb-nuclear interference effects. 
Nuclear couplings dominate only at the lowest partial waves. 
However, their contributions are small due to the factor $(2l+1)$ contained 
in the partial-wave expansion of the breakup cross section. \\

The fusion process is quite different. 
At high angular momenta, the centrifugal barrier prevents the system 
from reaching the strong absorption region, where the fusion processes 
occur. 
Thus, it is unsurprising that Coulomb couplings play a major role in the 
breakup process, but they have a negligible influence on the CF cross section.
Then, only $l\lesssim l_{\rm g}$ contributes to fusion, and the nuclear couplings are dominant at these angular 
momenta. \\

Now, we discuss the collisions of $^9$Be projectiles. 
They are also very interesting since their breakup threshold is only 1.65 MeV. 
There are available CF data for collisions of $^9$Be with several targets:  
$^{144}$Sm~\cite{GLP09}, $^{197}$Au~\cite{LWL19,KGT20,GAA21}, 
$^{186}$W~\cite{FGL13}, and $^{208}$Pb~\cite{DGH04}.
A comparative study of the CF data of these systems, reduced by the 
fusion function method, was presented in Fig.~13 of Ref.~\cite{GAA21}. 
This comparison was inconclusive since the data did not show any trend. 
The CF survival factors for systems in the same mass range differed considerably. 
However, we emphasize that the data for different targets were measured 
by different groups using different experimental techniques. 
Thus, a detailed study of the experimental procedures of each experiment 
would be needed before one can draw reliable conclusions and 
this would be beyond the scope of the present work. 
Nevertheless, the data for the $^9$Be + $^{208}$Pb and the 
$^{6,7}$Li + $^{209}$Bi systems, analyzed in the previous two figures,  
were measured by the same group using similar 
experimental techniques. 
Then, we restrict our study to $^9$Be + $^{208}$Pb. \\

The experimental CFF for the $^9$Be + $^{208}$Pb system is shown in 
Fig.~\ref{all-wb}(c), in comparison to the classical fusion line. 
One finds that the reduced data points are also distributed along the 
linear function of Eq.~(\ref{linear}), but now the CF survival factor is 
\begin{equation}
\alpha^{^9{\rm Be}} = 0.65.
\end{equation}
Note that ita value is intermediate between $\alpha^{^6{\rm Li}}$ and 
$\alpha^{^7{\rm Li}}$. 
Since the breakup threshold of this nucleus lies between the ones of the two 
Li isotopes, this result is consistent with the assumption that the CF survival factor decreases with the breakup threshold of the projectile.\\

Finally, we look at collisions of the two-neutron halo nucleus $^6$He. 
Since experiments involving unstable beams are much more challenging, fusion data 
involving $^6$He projectiles are very scarce. 
Nevertheless, there are good CF data for the $^6$He + $^{209}$Bi system at 
several collision energies and the  $^6$He + $^{238}$U system at a few 
collision energies. 
In typical situations, CF followed by the evaporation of two neutrons cannot 
be distinguished from ICF events with the capture of the $^4$He cluster. 
However, the CF cross sections could be measured under special conditions 
of these experiments. 
Kolata {\it et al.}~\cite{KGP98a,KGP98b} measured the CF cross section 
for the $^6$Li + $^{209}$Bi system at near-barrier energies. 
The CF of this system leads to the formation of the $^{215}$At, which 
evaporates 1, 2, 3, and 4 neutrons. 
The cross section was determined by measuring the characteristic alpha 
particles emitted by the $^{212}$At and $^{211}$At residual nuclei, which, 
according to statistical model code predictions~~\cite{KGP98a,KGP98b}, are the dominant decay modes within the energy
interval of the experiment. 
The $^{212}$At and $^{211}$At are also formed by the breakup of the projectile, 
followed by the capture of the $^4$He cluster (ICF$\alpha$). 
However, owing to excitation energy considerations~\cite{FRL23}, these nuclei 
could not emit the alpha particles detected in the experiment. 
Nevertheless, they are expected to make a small
contribution to the data points at the highest collision energies. \\

The CF cross section for the $^6$He + $^{238}$U system was measured by 
Raabe {\it et al.}~\cite{RSC04}. 
In this case, the signature of CF events was the detection of two fission 
fragments emitted back to back, unaccompanied by a third charged fragment 
with a projectile-like kinematic. 
In fact, the same residual nucleus can be formed by the CF and the 
ICF$\alpha$ processes after the evaporation of different numbers of neutrons 
(two more neutrons in the case of CF). 
However, comparing the fission barrier with the excitation 
energies in the CF and ICF$\alpha$ processes, one concludes that the 
latter does not contribute to the measured cross section~\cite{FRL23}.\\

Fig.~\ref{all-wb}(d)  shows the experimental CFFs corresponding to the 
CF data of the $^6$He + $^{209}$Bi and $^6$He + $^{238}$U systems. 
One observes that the results exhibit the same behavior as the CFFs for 
$^7$Li projectiles. 
The data points of both systems are very close to the black dotted line, 
which corresponds to the function of Eq.~(\ref{alpha07}). 
Note that the experimental CFF tends to grow above the dotted 
line at the two data points with the highest energies. 
This is likely due to spurious contributions from ICF$\alpha$ events,
which, according to PACE estimates, become relevant at the highest energies of 
the experiment.\\

In the calculation of the CFFs of Figs.~\ref{all-wb}(a), \ref{all-wb}(b), 
\ref{all-wb}(c), and \ref{all-wb}(d), we used the radii of the Coulomb 
barriers of the SPP and the approximate expression for $f_{\rm R}(y)$ 
(Eq.~\ref{ffit}). 
The barrier parameters of the SPP for the weakly bound systems 
studied in the present work are listed in Table~\ref{barpar-supfactor}.
\begin{table}%[h!]
\centering
\caption{Barrier parameters of the SPP for the weakly bound studied systems of
in Fig.~\ref{all-wb}.
}
\vspace{0.5cm}
\begin{tabular}{lcccc}
\hline 
 System  & \qquad \ \ \  $R_{\rm B}$ (fm)  \qquad  &\ \ \qquad $V_{\rm B}$ (fm) & \qquad $\hbar\omega$ (MeV)  \\ 
\hline  
$^4$He    + $^{209}$Bi   & \ \ \ \ \qquad 10.6   &\qquad 21.3 &\qquad 4.7   \\
$^{12}$C + $^{194}$Pt    & \ \ \ \ \qquad 11.3   &\qquad 55.6 &\qquad 4.1   \\
$^{12}$C + $^{208}$Pb    & \ \ \ \ \qquad 11.5   &\qquad 57.7 &\qquad 4.1   \\
$^{16}$O + $^{144}$Sm    & \ \ \ \ \qquad 10.8   &\qquad 61.4 &\qquad 3.9   \\
   &  &  &  \\
$^6$Li + $^{90}$Zr      & \ \ \ \ \qquad  9.6   &\qquad 16.6   & \qquad 3.4 \\
$^6$Li + $^{124}$Sn     & \ \ \ \ \qquad  10.3  &\qquad 19.5   & \qquad 3.6 \\
$^6$Li + $^{197}$Au     & \ \ \ \ \qquad  11.1  &\qquad 28.7   & \qquad 4.1 \\
$^6$Li + $^{198}$Pt     & \ \ \ \ \qquad  11.1  &\qquad 28.3   & \qquad 4.1 \\
$^6$Li + $^{209}$Bi     & \ \ \ \ \qquad  11.2  &\qquad 29.8   & \qquad 4.2 \\
   &  &  &  \\
$^7$Li + $^{124}$Sn    & \ \ \ \ \qquad  10.4 &\qquad 19.3   & \qquad 3.3  \\
$^7$Li + $^{197}$Au    & \ \ \ \ \qquad  11.3 &\qquad 28.3   & \qquad 3.8  \\
$^7$Li + $^{198}$Pt    & \ \ \ \ \qquad  11.3 &\qquad 27.9   & \qquad 3.7  \\
$^7$Li + $^{209}$Bi    & \ \ \ \ \qquad  11.4 &\qquad 29.4   & \qquad 3.8  \\
   &  &  &  \\
$^9$Be + $^{208}$Pb    & \ \ \ \ \qquad  11.5 &\qquad 38.5   & \qquad 3.9  \\
    &  &  & \\
$^6$He + $^{209}$Bi    & \ \ \ \ \qquad  11.6 &\qquad 16.4   & \qquad 3.4  \\
$^6$He  + $^{238}$U    & \ \ \ \ \qquad  11.9 &\qquad 21.0   & \qquad 3.4 \\
\hline
\end{tabular}
\label{barpar-supfactor}
\end{table}
%

%%%%%%%%%%%%%%%%%%%%%%%%%%%%%%%%%%%%%%%%%%%%%%%%%%%%%%%%%%%%%%%
\section{Conclusions}
\label{concl}

%%%%%%%%%%%%%%%%%%%%%%%%%%%%%%%%%%%%%%%%%%%%%%%%%%%%%%%%%%%

We derived an improved version of the Wong formula for heavy-ion fusion 
that is valid under conditions where the standard Wong formula is inaccurate. 
This is achieved by replacing the s-wave barrier parameters in the Wong formula 
with the parameters of the $l$-dependent potential barrier at an effective 
partial wave, given by an average over angular momentum. 
The same procedure is used to obtain an improved expression for
the classical fusion cross section. \\

The improved Wong cross sections for systems over a broad mass range were 
compared to the corresponding cross sections of the barrier penetration model. 
They were shown to be in excellent agreement. 
An analogous comparison was made for the improved classical fusion 
cross sections of the same systems. 
In this case, it was restricted to energies above the Coulomb barrier. 
The agreement with the barrier penetration model 
cross section was also extremely good.\\

Based on the improved classical cross sections, we proposed a new procedure 
to reduce the fusion data of different systems. 
This procedure, which we called the {\it classical fusion function} method, 
leads to a new universal function, the {\it classical fusion line}, which 
plays the role of a benchmark in comparisons of reduced fusion data. 
This method was used to study the suppression of complete fusion 
in collisions of weakly bound projectiles. 
Comparisons of reduced cross sections in collisions
of $^6$Li and $^7$Li with targets over a wide mass range lead to two 
conclusions: ({\it i}) The suppression increases as the breakup threshold 
of the projectile increases, ({\it ii}) the suppression 
does not seem to depend on the target charge. 
This indicates that the suppression of complete fusion 
is mainly due to nuclear breakup couplings.\\

%We also study the suppression of complete fusion for some reactions induced by $^9$Be and $^6$He. Due to the lack of considerable statistics for reactions induced by these two projectiles, we will not include them in our conclusions. More experimental data is needed to reach a definitive conclusion.\\ 

\section*{Acknowledgments}

Work supported in part by the Brazilian funding agencies, CNPq, FAPERJ, 
CAPES, and the INCT-FNA (Instituto Nacional de Ci\^encia e Tecnologia- 
F\'\i sica Nuclear e Aplica\c c\~oes), research project 464898/2014-5, 
and the Uruguayan agencies PEDECIBA (Programa de Desarrollo de las Ciencias 
B\'asicas) and ANII (Agencia Nacional de Innovaci\'on e Investigaci\'on).
We are indebted to Dr. Roberto Linares for critically reading the manuscript.

\bibliographystyle{apsrev}
%\bibliography{fusbreak-2023_vs03} 

\end{document}